\documentclass[usenatbib]{mnras}
\usepackage{newtxtext,newtxmath}

\usepackage[T1]{fontenc}
\usepackage{ae,aecompl}

\usepackage{multicol,graphicx,longtable,float,adjustbox,rotating,pdflscape,geometry,wrapfig,subfigure,threeparttablex,booktabs}
\graphicspath{{images/}}
\usepackage{amsmath}

\title[]{Supernova Remnants in M33: X-ray Properties as Observed by {\it XMM-Newton}}

\author[K. Garofali et al.]{Kristen Garofali,$^{1}$\thanks{E-mail: garofali@uw.edu (KG)}
 Benjamin F. Williams,$^{1}$
Paul P. Plucinsky,$^{4}$
Terrance J. Gaetz,$^{4}$
\newauthor Brian Wold,$^{1}$
Frank Haberl$^{2}$
Knox S. Long,$^{5}$
William P. Blair,$^{3}$
\newauthor Thomas G. Pannuti,$^{6}$
P. Frank Winkler,$^{7}$
Jacob Gross,$^{1}$
\\
$^{1}$University of Washington Astronomy Department, Box 351580, Seattle, WA, 98195 \\
$^{2}$Max-Planck-Institut f\"ur extraterrestrische Physik, Giessenbachstra{\ss}e, 85748 Garching, Germany \\
$^{3}$Department of Physics and Astronomy, Johns Hopkins University, 3400 North Charles Street, Baltimore, MD 21218 \\
$^{4}$Harvard-Smithsonian Center for Astrophysics, 60 Garden Street, Cambridge, MA 02138 \\
$^{5}$Space Telescope Science Institute, 3700 San Martin Drive, Baltimore, MD 21218 \\
$^{6}$Space Science Center, Department of Earth and Space Sciences, 235 Martindale Drive, Morehead State University, Morehead, KY 40351 \\
$^{7}$Department of Physics, Middlebury College, Middlebury, VT 05753
}

\pubyear{2017}

\begin{document}
\label{firstpage}
\pagerange{\pageref{firstpage}--\pageref{lastpage}}
\maketitle

\begin{abstract}

We have carried out a study of the X-ray properties of the supernova remnant (SNR) population in M33 with {\it XMM-Newton}, comprising deep observations of 8 fields in M33 covering all of the area within the D$_{25}$ contours, and with a typical luminosity of 7.1$\times$10$^{34}$ erg~s$^{-1}$ (0.2-2.0 keV) . Here we report our work to characterize the X-ray properties of the previously identified SNRs in M33, as well as our search for new X-ray detected SNRs.
With our deep observations and large field of view we have detected 105 SNRs at the 3$\sigma$ level, of which 54 SNRs are newly detected in X-rays, and three are newly discovered SNRs. Combining
 {\it XMM-Newton} data with deep {\it Chandra} survey data
allows detailed spectral fitting of 15 SNRs, for which we have measured temperatures, ionization timescales, and individual abundances. This large sample of SNRs allows us to construct an X-ray luminosity function, and compare its shape to luminosity functions from host galaxies of differing metallicities and star formation rates to look for environmental effects on SNR properties. We conclude that while metallicity may play a role in SNR population characteristics, differing star formation histories on short timescales, and small-scale environmental effects appear to cause more significant differences between X-ray luminosity distributions. In addition, we analyze the X-ray detectability of SNRs, and find that in M33 SNRs with higher [SII]/H$\alpha$ ratios, as well as those with smaller galactocentric distances, are more detectable in X-rays.

\end{abstract}


\section{Introduction}
Supernova remnants (SNRs) 
deposit energy and metals into the interstellar medium (ISM), and thus are key drivers of galactic chemical evolution. The ejecta and shock wave from the supernova (SN) explosion
interact with the ISM, making it visible in optical, radio
and X-ray wavelengths. The supernova events themselves are short-lived, so very few are available to study nearby in detail. The SNRs they leave behind, by contrast, offer a way to unlock information about the progenitor and its lasting impacts on the ISM. Typically, thermal X-ray spectra 
have been used to infer properties of the young, ejecta-dominated SNRs that are indicative of the supernova progenitor, and therefore significant 
effort has been put into X-ray detections \citep[e.g.][]{Vink2003,Badenes2003,Gaetz2007, Badenes2007, Reynolds2008,LongSNR,Vink2012,Yamaguchi2014, Pannuti2014,Long2014,Maggi2016}. 

It is difficult to systematically compare SNR properties, such as progenitor type, with their effect on the ISM, because most SNR studies in the Milky Way to date have focused on individual SNRs. In addition, SNR population studies in the Milky Way are difficult due to distance uncertainties and variable absorption to the individual objects
\citep{Woltjer1972,Milne1979,Raymond84,Green2014}. SNRs in Local Group galaxies, however, are in the unique position to be studied as a population in relation to the surrounding stellar population and ISM, because they are all at a common distance and have similar foreground extinction. In the Magellenic Clouds, the
progenitor types, spectral properties, X-ray morphologies, explosion types and size distribution of the SNR population have all been well characterized in optical, X-ray, and radio wavelengths \citep{Mathewson1973, Long1979,Mathewson1983,Chu1988,Hughes1998,Badenes2010, Lopez2011,Maggi2016,Maggi2014}. M31 hosts a population of optically identified SNRs \citep{Blair1981,Braun1993,Magnier1995,LeeLeeM31} for which progenitor masses have been estimated \citep{Jennings2012,Jennings2014}, and X-ray measurements have been made with {\it XMM-Newton} \citep{Sasaki2012}. Outside the Local Group, populations of extragalactic SNRs have been identified in nearby spiral galaxies in the optical, based on emission-line ratios \citep{Matonick1997,Blair2012, Leonidaki2013},  X-ray emission \citep{Pannuti2007,Leonidaki2010}, and based on radio observations \citep{Lacey2001,Pannuti2002}. 

M33 has perhaps the best-characterized SNR population of any spiral galaxy \citep{Sabbadin1979,Dodorico1980,Gordon1998,LongSNR,Sarbad2017}, making it a prime target to extend these previous works characterizing extragalactic SNRs. In particular, M33, a late-type Sc spiral, is well-suited for X-ray studies of SNRs
because of its proximity to the Milky Way at 817$\pm$58~kpc
\citep{FreedmanDistance}, its close to face-on angle of inclination,
\textit{i} = 56$^{\circ}\pm$1$^{\circ}$ \citep{ZaritskyAngle}, and its low foreground absorption \citep[N$_{H}$$\approx$6$\times$10$^{20}$cm$^{-2}$,][]{StarkAbsorption}.
Previous detailed multi-wavelength surveys have revealed a rich SNR population
(218 candidates, 86 confirmed via multi-wavelength detections) in optical, radio, and X-ray
wavelengths. Using optical emission-line ratios \citet{Gordon1998} identified a population of 98 SNR candidates in M33. Recently, \citet[][hereafter L10]{LongSNR} carried out a multiwavelength study of 131 of the previously known 137 SNRs in the galaxy; they detected 82 (58) at the 2$\sigma$ (3$\sigma$) level with {\it Chandra}, and obtained upper limits for the rest. Most recently, \citet[][hereafter LL14]{LeeLeeM33} boosted the number to 199 optically selected SNR candidates, of which 78 were not previously reported in L10's catalog. Here we carry out an analysis of the properties of all 218 known and suggested SNRs in M33.  This includes the 137 sources described by L10 (of which 121 overlap with the sources discussed by LL14), the 78 new sources identified as candidates by LL14, and 3 X-ray candidates identified by \citet[][hereafter W15]{Williams2015}.

Herein we utilize data from a deep survey of M33 using an 8 field {\it XMM-Newton} mosaic that extends out to the D$_{25}$ isophote. The point source catalog from this survey was published by W15. In this paper, we leverage both the excellent soft sensitivity of {\it XMM-Newton}, as well as the large field-of-view from the W15 survey as they pertain to the SNR population in M33. With this \textit{XMM-Newton} survey, all 218 SNR candidates are within the field-of-view and refinements can be made to the
properties of those SNRs already detected at X-ray and/or optical wavelengths. For the purposes of obtaining X-ray spectral fits we have also made use of {\it Chandra} ACIS Survey of M33 \citep[ChASeM33,][]{Tullmann} data where possible. In section 2, we describe the data used from this and 
previous surveys as well as the 
data reprocessing and reduction techniques. Section 3 outlines the SNR catalog, and details the characterization of the SNR population based on spectral fitting, hardness ratios (HRs), and X-ray morphology. In Section 4 we discuss the results from this 
most recent X-ray survey of the M33 SNR population, including the shape of the X-ray luminosity function, and the implications for SNR detectability, and finally in Section 5 we present our conclusions. 

\section{Survey Overview} \label{sec:surveydat}

The observations and analysis of the {\it XMM-Newton} survey data of M33 used in this paper are
described by W15. The W15 survey consists of an 8 field {\it XMM-Newton} mosaic of M33 with a summed exposure time of 900 ks extending out to the D$_{25}$ isophote and to a limiting 0.2-4.5 keV luminosity of 4$\times$ 10$^{34}$ erg~s$^{-1}$ at the distance of M33. For the purpose of analyzing the SNR population, we have optimized our reduction of the survey data for 
extended sources, as we describe in Section~\ref{sec:reprocessing} and in the Appendix. All 218 previously identified SNR candidates are within the field-of-view of W15, allowing for cross-correlation of the W15 X-ray catalog with both the existing X-ray and optical catalogs of L10 and LL14, as well as identification of 3 new X-ray selected SNRs (described in Section~\ref{sec:cat}) based on X-ray HRs and visual inspection of Local Group Galaxy Survey (LGGS) data \citep{Massey2006}. The positions for all sources used in the remainder of this analysis come from the catalog of W15. The X-ray fluxes, and HRs reported in Table~\ref{tab:sourcetab} and Table~\ref{tab:3sigtab} come from custom measurements at the positions of all 218 SNR candidates using the {\it XMM-Newton} data in the bands described in Section~\ref{sec:reprocessing}. The {\it XMM-Newton} spectra used in the spectral fitting described in Section~\ref{sec:spectralfit} were extracted specifically for this analysis as described in the Appendix.

In addition to the catalog of W15, we utilize high resolution observations
from the {\it Chandra} ACIS Survey of M33 \citep[ChASeM33,][]{Tullmann}, which had a total exposure time of 1.4 Ms and covered about 70\% of the D$_{25}$ isophote down to a limiting 0.35-8.0 keV luminosity of 2.4$\times$10$^{34}$ ergs~s$^{-1}$. The SNR catalog from the ChASeM33 survey is
described in L10, and, in addition to cross-correlating 
our SNR candidates with those of L10, we also use their {\tt ACIS} spectra when available for spectral fits. 

We also cross-correlate our sources with the optically selected SNR candidate catalog of LL14. Their survey used narrow-band images from the LGGS \citep{Massey2006} to identify SNR candidates based on emission-line ratios ([SII]/H$\alpha$ $\textgreater$ 0.4) and shell-like or circular morphology for sources smaller than 100~pc. The L10 survey, by contrast, did not initially cut candidates based on morphological or size considerations, focusing instead on evidence of shock-heating, and further used only portions of the LGGS within the {\it Chandra} footprint. For the purposes of this paper, we have kept all objects contained in both survey lists. We discuss our measurements at the locations of all sources in each catalog in Section~\ref{sec:cat}.

In addition to finding SNR candidates, LL14 assign each candidate in their catalog a tentative progenitor classification of core-collapse (CC) or Type Ia based upon the surrounding 
stellar population. CC SNe result from the explosion 
of a massive star and thus are expected to be in regions of recent star formation nearby other OB stars. Type Ia, or thermonuclear, SNe are caused by the detonation of a white dwarf (WD) that has 
reached its Chandrasekhar limit and are expected in regions of little to no recent star formation (i.e. few nearby OB stars). However, it is possible a Type Ia SN could occur near a region of OB stars in areas with multiple epochs of star formation, highlighting the need for a full star formation history of the surrounding region to determine more reliably the progenitor class. Such classifications exist for a handful of SNRs in M33 from the work of \citet{Jennings2014}. These SNR progenitor classifications are the result of measured progenitor masses from detailed star formation histories of the stellar populations surrounding each SNR, and are considered robust determinations. When the classifications from \citet{Jennings2014} are unavailable, we instead use the LL14 tentative progenitor type labels. Together these classifications are used to explore the efficacy of measured 
HRs (Section~\ref{sec:hr}) in typing SNR progenitors. 

\subsection{Reprocessing of \textit{XMM-Newton} Data}\label{sec:reprocessing}

While the vast majority of the reduction techniques for the analysis
presented here was described in detail in W15, there were some special
considerations that we employed for source detection and background
characterization when looking specifically for the SNRs in M33, which
slightly differ from the description in W15. 

For detecting and measuring extended sources, it is beneficial to use the software provided by
the \texttt{Extended Source Analysis Software (ESAS)} (Kuntz \& Snowden 2008), 
a package within \texttt{SAS} optimized for extended sources,  which models the
background light curves during an observation and identifies time periods 
when the background level is significantly enhanced compared to the quiescent 
background, allowing cleaner separation of extended sources from the
background as well as spectra with less background contamination.

We selected good-time intervals (GTIs) using the \texttt{ESAS} tools \texttt{pn-filter} and \texttt{mos-filter}. Utilizing the unexposed corner
sections of the CCDs, count thresholds are chosen by fitting a Gaussian peak at the quiescent
count rate. A GTI file is produced which includes only time intervals where the count rate was
within 2$\sigma$ of the peak of the aforementioned Gaussian. Within the \texttt{ESAS} package \texttt{pn-filter} and \texttt{mos-filter} were applied to the \texttt{SAS} task
\texttt{espfilt} to determine the GTIs. We then applied these GTIs to our event lists for spectral extraction. Our spectral extraction required a large amount of 
customization. We therefore place a detailed explanation in the Appendix. The extracted spectra are used 
for the spectral fitting described in Section~\ref{sec:simulfit}.

When searching for soft and extended gas emission sources (such as large SNRs or HII regions),
we applied the same \texttt{emosaicprep} and \texttt{emosaicproc} tasks as for the point sources, but using the
\texttt{ESAS}-processed event lists. For the purposes of
detecting SNRs, the energy range was set to 0.2-2.0 keV and the positions of the L10 and LL14 SNR candidates were input
into {\tt emldetect}. For the 0.2-2.0~keV band we ran {\tt emldetect} simultaneously on the 0.2-0.5~keV, 0.5-1.0~keV, and 1.0-2.0~keV bands, and let {\tt emldetect} calculate the full band (0.2-2.0~keV) totals based on the provided exposure times, background images, and masks. This energy range was chosen to leverage the soft sensitivity of {\it XMM-Newton} for SNR detection. The pn requires more conservative flagging be applied to the event list in the softest band (0.2-0.5~keV) to avoid spurious detections. For this reason, the individual bands were run separately with flagging as described in W15, and then combined to create a full band (0.2-2.0~keV). For the
purposes of measuring and comparing luminosities, the energy range was also set to
0.35-2.0 keV to match the band of L10.  These measurements at the locations of previously known SNRs are discussed in \ref{sec:cat}.

Another critical adjustment to
make was the choice of energy conversion factor (ECF) values. The list of source positions from {\tt emosaicproc} was fed to {\tt emldetect} 
to calculate on-axis equivalent count rates and convert these values to a flux using ECFs selected
during the detection script. Table~\ref{tab:ecftab} lists the ECFs used to calculate the fluxes in the 0.2-2.0 keV
band (plus component bands) which was used for source detection, and the
0.35-2.0 keV band  which was used for comparison to L10, and the 0.3-0.7 keV, 
0.7-1.1. keV, and 1.1-4.2 keV bands, which were used for HR 
calculations in Section~\ref{sec:hr}. The unabsorbed ECFs were calculated based on \texttt{XSPEC} simulations of an \texttt{apec} spectrum with absorption with parameters N$_{\rm H}$=1x10$^{21}$~cm$^{-2}$, kT = 0.6 keV, and elemental abundance set to half solar. 
This spectrum was also chosen to remain consistent with L10.
Futhermore, we ensured that the locations of all SNR candidates were
included in the candidate source list that was measured by the
{\tt emldetect} step of our analysis routine.  Thus, we were able to obtain either detections or
upper limits for all SNR candidates. Occasionally, {\tt emldetect} fails to properly combine the individual bands to produce a reliable full band total. In these cases (8\% of sources), the source counts in the full band (0.2-2.0 keV) are a factor of 2 discrepant from the sum total of the individual bands (0.2-0.5 keV, 0.5-1.0 keV, and 1.0-2.0 keV ). We denote these sources with a `t' flag in Table~\ref{tab:sourcetab} and Table~\ref{tab:3sigtab}. These sources still have reliable measurements in the individual bands, so for their full band 0.2-2.0~keV totals we report the counts, count rate, and flux values as the sum of the individual bands. For all other sources the total flux values are output from {\tt emldetect} and represent the sum of the fluxes from each EPIC instrument weighted by the appropriate calibration files. All sources are listed in Table~\ref{tab:sourcetab}. 

\section{Results}\label{sec:results}

Our goal is to provide the best possible characterization of the SNR population of M33 given all of the available data. Because {\it XMM-Newton} provides high
soft-band sensitivity, this new survey provides further constraints on SNR spectral fits and HR measures which can potentially be used to constrain progenitor explosion type. We describe the sample of SNRs measured in this survey, as well as each of these methods as applied to that sample below.  

\subsection{Catalog of M33 SNRs}\label{sec:cat}

\begin{table}
\centering
\caption{Unabsorbed energy correction factors (ECFs) for the bands and instruments used in this survey. ECF units are counts~erg~cm$^{-2}$.}
\label{tab:ecftab}
\begin{tabular}{cccc}
\hline \hline
Energy Band (keV) & pn & MOS1 & MOS2 \\
\hline
0.35-2.0 & 1.8$\times$10$^{-12}$ & 7.45$\times$10$^{-12}$ & 7.44$\times$10$^{-12}$ \\
0.2-2.0 & 2.5$\times$10$^{-12}$ & 8.2$\times$10$^{-12}$ & 8.1$\times$10$^{-12}$ \\
0.2-0.5 & 4.2$\times$10$^{-12}$ &  2.7$\times$10$^{-11}$ & 2.8$\times$10$^{-11}$ \\ 
0.5-1.0 & 1.8$\times$10$^{-12}$ & 7.8$\times$10$^{-12}$ & 7.8$\times$10$^{-11}$ \\
1.0-2.0 & 1.9$\times$10$^{-12}$ & 5.9$\times$10$^{-12}$ & 5.8$\times$10$^{-12}$ \\
0.3-0.7 & 1.8$\times$10$^{-12}$ & 1.1$\times$10$^{-11}$ & 1.1$\times$10$^{-11}$ \\
0.7-1.1 & 1.9$\times$10$^{-12}$ & 7.8$\times$10$^{-12}$ & 7.7$\times$10$^{-12}$ \\
1.1-4.2 & 2.0$\times$10$^{-12}$ & 6.1$\times$10$^{-12}$ & 6.1$\times$10$^{-12}$ \\
\hline
\end{tabular}
\end{table}

To measure fluxes or, where not possible, establish upper limits on the flux of SNRs in M33, we measured the locations of SNR candidates from L10, LL14, and W15 (218 total sources) 
in both the 0.2-2.0~keV band as well as the 0.35-2.0~keV band to be consistent with L10. We find that the average signal-to-noise is higher in the 0.2-2.0~keV band, and after verifying all measurements by eye to remove spurious detections in both bands, we found that using the 0.2-2.0~keV band results in 12 more 3$\sigma$ detections than using the 0.35-2.0~keV band. We therefore conclude that the 0.2-2.0~keV band is better for detection of SNRs when using {\it XMM-Newton} and that only 3$\sigma$ measurements should be taken as reliable detections, as there can be fluctuations in the background at the 2$\sigma$ level in the 0.35-2.0~keV band on the size scale of SNRs in the {\it XMM-Newton} imaging. 


We have also inspected all measurements at the locations of SNR candidates by eye to validate the detections and non-detections in our sample. In some cases, overlapping sources in the 
{\it XMM-Newton} data lead to erroneously high fluxes for a single SNR candidate. For these sources, if there is a previous X-ray detection of the SNR (i.e. from L10), the source is flagged as contaminated by the overlapping source (`c'), and the measured flux is likely too high and thus treated as an upper limit. If there is a nearby contaminating X-ray source, but no previous X-ray detection at the location of the SNR candidate, the source is denoted with the `x' flag, and the flux is treated as an upper limit, and the source a non-detection. Those sources that did not appear to be reliable detections in the by-eye validation are denoted by the `n' flag, and their fluxes are also reported as upper limits, and the sources treated as non-detections. As noted in Section~\ref{sec:reprocessing} sources marked `t' have full band totals that come from the sum of the individual band runs from {\tt emldetect}. Sources with erroneous {\tt emldetect} count errors in at least one band are denoted by the `e' flag, and their count errors are pegged to the total number of counts in that band. 

All 218 sources are recorded in Table~\ref{tab:sourcetab}. We report ID numbers in this catalog, corresponding ID numbers in both L10 and LL14, {\it XMM-Newton} positions, counts in the 0.2-2.0~keV band, count rates and associated errors in the 0.2-2.0~keV band in s$^{-1}$, fluxes from {\tt elmdetect} (sum of all EPIC instruments) in the 0.2-2.0~keV band in ergs~cm$^{-2}$~s$^{-1}$ (used for detection), fluxes in the 0.35-2.0~keV band in ergs~cm$^{-2}$~s$^{-1}$ (used for comparison to L10), SNR sizes (in pc) from L10 and LL14, [SII]/H$\alpha$ from both L10 and LL14, and the log of H$\alpha$ luminosity in ergs~s$^{-1}$ from both L10 and LL14. Individual source ID numbers are denoted with the `c', `x', `n', `t', and `e' flags as described above. Sources that are upper limit measurements in this survey have fluxes preceded by $\textless$. 

In addition to listing all 218 sources in Table~\ref{tab:sourcetab}, we also list all 3$\sigma$ detections only in Table~\ref{tab:3sigtab}, along with their associated counts (sum of all EPIC instruments) and count errors in the 0.2-2.0~keV, 0.35-2.0~keV, 0.3-0.7~keV, 0.7-1.1~keV, and 1.1-4.2~keV bands. The last three bands are used to compute HRs for all detected sources only as described in Section~\ref{sec:hr}. The HRs in these bands, computed with counts using the
Bayesian Estimation of Hardness Ratios (BEHR) method \citet{Park2006}, are listed with their associated errors in columns 9-10 of Table~\ref{tab:3sigtab}. Columns 11-12 similarly list the HRs and associated errors calculated from the 0.2-0.5~keV, 0.5-1.0~keV, and 1.0-2.0~keV bands using BEHR. The final column of Table~\ref{tab:3sigtab} denotes the level at which the source was measured in the 0.35-2.0~keV band in L10: $\textless$ for upper limit, 2$\sigma$, or 3$\sigma$. The comparison of this catalog with those of L10 and LL14 is described below. 

The vast majority of extragalactic SNRs, including those in M33, were first identified optically based on elevated [SII]/H$\alpha$ ratios as compared to HII regions. This technique works well in general, especially for brighter objects, but HII contamination for fainter objects, especially in complex regions, can affect the observed [SII]/H$\alpha$ ratio and cause uncertainty in some optical identifications. We therefore view 3$\sigma$ X-ray detection of a previously optically identified SNR to be a strong confirmation of SNR detection based on both the elevated [SII]/H$\alpha$ as well as strong X-ray emission. Of course, some of the optically identified SNR candidates that are undetected in X-rays may simply fall below our detection threshold and thus still be SNR detections. However, for the reasons described here, and in Section~\ref{sec:cat}, we consider sources that are identified optically as well as measured at 3$\sigma$ confidence in X-rays to be well-confirmed SNRs. These sources, and those detected in X-rays at the 3$\sigma$ level from L10, are regarded as X-ray confirmed SNRs in all subsequent detectability analyses, while any candidates measured less than 3$\sigma$ are considered non-detections in the analysis that follows.  

We cross-correlate all our SNR candidates with measurements from LL14 and L10 to determine the number of newly X-ray detected SNRs in this catalog. The position of each of the 137 L10 sources was inspected by eye for a counterpart not already known to be a point source in the W15 catalog. To cross-correlate with LL14, we searched for counterparts in the W15 catalog out to a maximum separation of 10" for those sources not already matched to sources in L10. This resulted in 69 matches of the 78 newly reported sources in LL14. The remainder of the LL14 sources (121) were previously matched to counterparts in L10. For those sources in both L10 and LL14 that did not have a counterpart in W15 after cross-correlation we forced {\tt emldetect} to make measurements at the locations of these sources to ensure that we would measure upper limits for all sources.

\begin{landscape}
\setlength\LTcapwidth{\linewidth}
\onecolumn
\small
\setlength\LTleft{-3in}
\setlength\LTright{-3in}
\begin{ThreePartTable}
\begin{TableNotes}
\footnotesize
\item [c] Flux is contaminated by nearby bright X-ray source, but source has a previous X-ray detection. Flux is an upper limit. 
\item [b] Unrealistic source count errors from {\tt emldetect} in at least one individual band. Source count errors pegged to total count values.
\item [n] Determined to be a nondetection via by-eye catalog checking. Flux is an upper limit.
\item [t] Total counts in the full band are discrepant by a factor of two from the summation of counts in each individual band due an {\tt emldetect} merging issue. Total counts in the 0.2-2.0 keV band are reported as the sum of counts in the 0.2-0.5~keV, 0.5-1.0~keV, and 1.0-2.0~keV band.
\item [x] Flux is contaminated by nearby bright X-ray source, but source does not have a previous X-ray detection. Flux is an upper limit.
\end{TableNotes}
\begin{longtable}{|l|r|r|r|r|r|r|r|r|r|r|r|r|r|r|r|}
\caption{ All 218 SNR and SNR candidates in our sample. Columns 1-3: lD number from this catalog, ID number in L10, and ID number in LL14. Columns 4-5: source RA and Dec from this catalog. Columns 6-7: source counts and associated errors, and count rates and associated errors in the 0.2-2.0 keV band (used for detection). Columns 8-9: fluxes in the 0.2-2.0 keV band (used for detection), and fluxes in the 0.35-2.0 keV band (used for comparison to L10). Columns 10-11: optical SNR sizes from the L10 and LL14 catalogs. Columns 12-13: [SII]/H$\alpha$ ratio from the L10 and LL14 catalogs (used for optical selection of candidates). Columns 14-15: log of the H$\alpha$ luminosity from the L10 and LL14 catalogs. All units are given in the column headers.} \label{tab:sourcetab} \\
\hline 
\multicolumn{1}{|c|}{ID} &
   \multicolumn{1}{|c|}{L10} &
   \multicolumn{1}{|c|}{LL14} &
   \multicolumn{1}{|c|}{RA} &
   \multicolumn{1}{|c|}{Dec} &
   \multicolumn{1}{|c|}{0.2-2.0 Cts} &
   \multicolumn{1}{|c|}{0.2-2.0 Ct Rate (s$^{-1}$)} &
   \multicolumn{1}{|p{1.7cm}|}{\centering 0.2-2.0 Flux \\ (ergs~cm$^{-2}$~s$^{-1}$)} &
   \multicolumn{1}{|p{1.7 cm}|}{\centering 0.35-2.0 Flux \\ (ergs~cm$^{-2}$~s$^{-1}$)} &
   \multicolumn{1}{|p{1cm}|}{\centering D(pc) \\ L10} &
   \multicolumn{1}{|p{1cm}|}{\centering D(pc) \\ LL14} &
   \multicolumn{1}{|p{1 cm}|}{\centering [SII]/H$\alpha$ \\ L10} &
   \multicolumn{1}{|p{1 cm}|}{\centering [SII]/H$\alpha$ \\ LL14} &
    \multicolumn{1}{|p{1.3 cm}|}{\centering log(L(H$\alpha$) (ergs~s$^{-1}$))  \\ L10} &
    \multicolumn{1}{|p{1.3 cm}|}{\centering log(L(H$\alpha$)  (ergs~s$^{-1}$)) \\ LL14} \\
\hline
\endfirsthead
   \hline
   \multicolumn{1}{|c|}{ID} &
   \multicolumn{1}{|c|}{L10} &
   \multicolumn{1}{|c|}{LL14} &
   \multicolumn{1}{|c|}{RA} &
   \multicolumn{1}{|c|}{Dec} &
   \multicolumn{1}{|c|}{0.2-2.0 Cts} &
   \multicolumn{1}{|c|}{0.2-2.0 Ct Rate (s$^{-1}$)} &
   \multicolumn{1}{|p{1.7cm}|}{\centering 0.2-2.0 Flux \\ ergs~s~cm$^{-2}$} &
   \multicolumn{1}{|p{1.7cm}|}{\centering 0.35-2.0 Flux \\ ergs~s~cm$^{-2}$} &
   \multicolumn{1}{|p{1cm}|}{\centering D(pc) \\ L10} &
   \multicolumn{1}{|p{1cm}|}{\centering D(pc) \\ LL14} &
   \multicolumn{1}{|p{1 cm}|}{\centering [SII]/H$\alpha$ \\ L10} &
   \multicolumn{1}{|p{1 cm}|}{\centering [SII]/H$\alpha$ \\ LL14} &
    \multicolumn{1}{|p{1.3 cm}|}{\centering log(L(H$\alpha$) (ergs~s$^{-1}$)) \\ L10} &
    \multicolumn{1}{|p{1.3 cm}|}{\centering log(L(H$\alpha$) (ergs~s$^{-1}$))  \\ LL14} \\
\hline
\endhead

  \hline
\endfoot

  \hline
  \insertTableNotes  

\endlastfoot
XMM-001 & -- & 1 & 1:32:25.78 & 30:30:04.03 & 21 $\pm$ 11 & 2.34e-04 $\pm$ 1.30e-04 & <8.20e-16 & <2.18e-16 & -- & 81.0 & -- & 0.74 & -- & 36.16 \\
XMM-002 & -- & 2 & 1:32:27.85 & 30:35:44.58 & 25 $\pm$ 15 & 2.81e-04 $\pm$ 1.80e-04 & <6.77e-16 & <3.44e-16 & -- & 72.0 & -- & 0.98 & -- & 36.05 \\
XMM-003 & 1 & -- & 1:32:30.83 & 30:27:45.99 & 107 $\pm$ 19 & 1.15e-03 $\pm$ 2.09e-04 & 1.40e-15  & 1.12e-15  & 123.0 & -- & 0.77 & -- & 36.98 & -- \\
XMM-004 & 2 & 3 & 1:32:31.41 & 30:35:32.90 & 66 $\pm$ 19 & 7.76e-04 $\pm$ 2.49e-04 & 1.09e-15  & 8.62e-16 & 29.0 & 28.0 & 0.44 & 0.59 & 36.20 & 36.02 \\
XMM-005 & -- & 4 & 1:32:35.36 & 30:35:19.81 & 27 $\pm$ 14 & 2.82e-04 $\pm$ 1.51e-04 & <7.61e-16 & <2.23e-16 & -- & 85.0 & -- & 0.50 & -- & 36.36 \\
XMM-006 & -- & 5 & 1:32:37.36 & 30:18:03.10 & 296 $\pm$ 23 & 6.02e-03 $\pm$ 4.93e-04 & 8.49e-15  & 7.30e-15  & -- & 85.0 & -- & 0.60 & -- & 36.15 \\
XMM-007 & -- & 6 & 1:32:39.78 & 30:27:54.95 & 11 $\pm$ 13 & 1.07e-04 $\pm$ 1.19e-04 & <3.59e-16 & <1.76e-16 & -- & 36.0 & -- & 0.93 & -- & 35.36 \\
XMM-008 & -- & 7 & 1:32:40.23 & 30:16:21.24 & 4 $\pm$ 7 & 1.16e-04 $\pm$ 2.16e-04 & <5.33e-16 & <2.85e-16 & -- & 44.0 & -- & 0.52 & -- & 35.65 \\
\end{longtable}
\end{ThreePartTable}
\end{landscape}

\begin{landscape}
\setlength\LTcapwidth{\linewidth}
\onecolumn
\small
\setlength\LTleft{-3in}
\setlength\LTright{-3in}
\begin{ThreePartTable}
\begin{TableNotes}
\footnotesize
\item [e] Unrealistic source count errors from {\tt emldetect} in at least one individual band. Source count errors pegged to total count values.
\item [t] Total counts in the full band are discrepant by a factor of two from the summation of counts in each individual band due an {\tt emldetect} merging issue. Total counts in the 0.2-2.0 keV band are reported as the sum of counts in the 0.2-0.5~keV, 0.5-1.0~keV, and 1.0-2.0~keV band. 
\end{TableNotes}
\begin{longtable}{|l|r|r|r|r|r|r|r|r|r|r|r|r|r|}
\caption{All 105 3$\sigma$ SNRs detected in our sample. Columns 1-3: lD number from this catalog, and the RA and Dec from this catalog. Column 4: source counts and associated errors in the 0.2-2.0 keV band (used for detection). Column 5: source counts and associated errors in the 0.35-2.0 keV band (used for comparison to L10). Columns 6-8: source counts and errors in the 0.3-0.7 keV, 0.7-1.1 keV, and 1.1-4.2 keV bands, which are used for computing the HRs shown in Figure~\ref{fig:triangleHR_maggibands}. Columns 9-10: HRs and errors in the aforementioned bands (HR1 = (M-S)/(H+M+S), HR2 = (H-M)/(H+M+S)). Columns 11-12: HRs and errors computed from the 0.2-0.5 keV, 0.5-1.0 keV, and 1.0-2.0 keV bands (HR1$_{XMM}$ = (M-S)/(S+M), HR2$_{XMM}$ = (H-M)/(H+M)). Column 13: measurement level in L10: 3$\sigma$, 2$\sigma$, or upper limit ($\textless$).}
\label{tab:3sigtab} \\
\hline
   \multicolumn{1}{|c|}{ID} &
   \multicolumn{1}{|c|}{RA} &
   \multicolumn{1}{|c|}{Dec} &
   \multicolumn{1}{|c|}{0.2-2.0 keV Cts} &
   \multicolumn{1}{|c|}{0.35-2.0 keV Cts} &
   \multicolumn{1}{|c|}{0.3-0.7 keV Cts} &
   \multicolumn{1}{|c|}{0.7-1.1 keV Cts} & 
   \multicolumn{1}{|c|}{1.1-4.2 keV Cts} &
   \multicolumn{1}{|c|}{HR1} &
   \multicolumn{1}{|c|}{HR2} &
    \multicolumn{1}{|c|}{HR1$_{XMM}$} &
    \multicolumn{1}{|c|}{HR2$_{XMM}$} &
   \multicolumn{1}{|c|}{L10 Detect} \\

\hline
\endfirsthead
   \hline
    \multicolumn{1}{|c|}{ID} &
   \multicolumn{1}{|c|}{RA} &
   \multicolumn{1}{|c|}{Dec} &
   \multicolumn{1}{|c|}{0.2-2.0 keV Cts} &
   \multicolumn{1}{|c|}{0.35-2.0 keV Cts} &
   \multicolumn{1}{|c|}{0.3-0.7 keV Cts} &
   \multicolumn{1}{|c|}{0.7-1.1 keV Cts} & 
   \multicolumn{1}{|c|}{1.1-4.2 keV Cts} &
     \multicolumn{1}{|c|}{HR1} &
   \multicolumn{1}{|c|}{HR2} &
    \multicolumn{1}{|c|}{HR1$_{XMM}$} &
    \multicolumn{1}{|c|}{HR2$_{XMM}$} &
   \multicolumn{1}{|c|}{L10 Detect} \\
\hline
\endhead

  \hline
\endfoot

  \hline
    \insertTableNotes  

\endlastfoot
XMM-003 & 1:32:30.83 & 30:27:45.99 & 107 $\pm$ 19 & 98 $\pm$ 19 & 55 $\pm$ 13 & 48 $\pm$ 11 & 15 $\pm$ 11 & -0.14 $\pm$ 0.02 & -0.29 $\pm$ 0.01 & 0.49 $\pm$ 0.05 & 0.18 $\pm$ 0.04 & -- \\
XMM-004 & 1:32:31.41 & 30:35:32.90 & 66 $\pm$ 19 & 53 $\pm$ 18 & 10 $\pm$ 10 & 25 $\pm$ 9 & 13 $\pm$ 16 & 0.04 $\pm$ 0.02 & -0.34 $\pm$ 0.02 & -- & -- & < \\
XMM-006 & 1:32:37.36 & 30:18:03.10 & 296 $\pm$ 23 & 286 $\pm$ 22 & 45 $\pm$ 9 & 104 $\pm$ 12 & 222 $\pm$ 20 & -0.16 $\pm$ 0.02 & -0.31 $\pm$ 0.02 & 0.47 $\pm$ 0.01 & -0.47 $\pm$ 0.01 & -- \\
XMM-009 & 1:32:40.53 & 30:16:37.20 & 74 $\pm$ 13 & 79 $\pm$ 13 & 12 $\pm$ 6 & 14 $\pm$ 6 & 91 $\pm$ 14 & 0.10 $\pm$ 0.03 & -0.21 $\pm$ 0.03 & 0.90 $\pm$ 0.07 & 0.60 $\pm$ 0.03 & -- \\
XMM-015 & 1:32:46.61 & 30:34:37.00 & 253 $\pm$ 26 & 207 $\pm$ 25 & 163 $\pm$ 19 & 35 $\pm$ 10 & 27 $\pm$ 14 & -0.31 $\pm$ 0.05 & -0.16 $\pm$ 0.05 & 0.47 $\pm$ 0.22 & 0.79 $\pm$ 0.04 & 3$\sigma$ \\
XMM-019 & 1:32:53.36 & 30:48:23.06 & 53 $\pm$ 15 & 41 $\pm$ 14 & 18 $\pm$ 9 & 22 $\pm$ 8 & 8 $\pm$ 9 & 0.02 $\pm$ 0.06 & -0.33 $\pm$ 0.09 & -- & -- & 2$\sigma$ \\
XMM-022 & 1:32:55.96 & 30:40:33.57 & 134 $\pm$ 23 & 123 $\pm$ 23 & 59 $\pm$ 14 & 32 $\pm$ 10 & 61 $\pm$ 18 & -0.30 $\pm$ 0.07 & -0.05 $\pm$ 0.10 & 0.60 $\pm$ 0.06 & 0.27 $\pm$ 0.04 & < \\
XMM-023 & 1:32:56.12 & 30:33:30.44 & 68 $\pm$ 20 & 57 $\pm$ 18 & 32 $\pm$ 12 & 19 $\pm$ 10 & 23 $\pm$ 14 & -0.73 $\pm$ 0.05 & -0.01 $\pm$ 0.05 & -- & -- & -- \\
\end{longtable}
\end{ThreePartTable}
\end{landscape}
\twocolumn

Although LL14 report 79 new SNR candidates over previous works, we find one of these 79 sources to have a potential counterpart in L10 (XMM-189, L10-122, LL14-174). As a result, our census of SNR candidates totals 218: 137 from L10, 78 from LL14, and three from W15. From our measurements of the locations of these SNRs we detect 105 at 
3$\sigma$ confidence and 145 at 2$\sigma$ confidence. We measure upper-limits for the remaining 73 non-detections. Of our 105 3$\sigma$ detections, 54 are newly detected in X-rays at 3$\sigma$, 3 are newly discovered in X-rays from this data set, 48 are 3$\sigma$ detections in both L10 and this work, and 96 are reported in LL14. There are six SNRs that were detected in L10 but are not detected here. Of the six L10 detections that are undetected here, the majority have between 2-8 times more exposure in the ChASeM33 survey due to overlapping observations than the exposure times from W15. The other sources are on regions of the detector that are unfavorable to detection i.e. far off-axis, or near a chip gap on the detector.  There are 25 SNRs which are upper limits in both L10 and this work. Of the 78 SNR candidates in LL14 and not in L10, we measure 18 as 3$\sigma$ detections, 39 at 2$\sigma$ confidence, and the remaining 39 as upper limits. In general, we are less likely to detect in X-rays the SNR candidates newly reported in LL14 compared to those SNR candidates reported in both L10 and LL14. This is very likely because the new LL14 objects have a lower mean surface brightness than the SNRs previously reported in L10, indicating that they are older and/or interacting with less dense ISM, both of which tend toward lower expected X-ray emission.

The three newly discovered SNRs were first reported in W15. The brightest of these is denoted as XMM-034 here.
This SNR (source 383 in W15) is now the fifth brightest X-ray emitting SNR in M33 with a 0.35-2 keV flux of 2.77 $\times$ 10$^{-14}$ erg~cm$^{-2}$~s$^{-1}$ (L$_{x}$(0.35-2.0~keV) = 2.2 $\times$ 10$^{36}$~erg~s$^{-1}$). The discovery of these new SNRs was facilitated by the larger survey area and increased soft sensitivity of {\it XMM-Newton}. In particular, any source that had strong emission below 1 keV compared to
above 1 keV and was not already classified as an SNR was studied in
the [S II] and H$\alpha$ images of \citet{Massey2006} to see if the
region hosted an SNR. This method for SNR candidate detection is discussed in more detail in W15. 

\begin{figure}
\centering
\includegraphics[scale=0.6]{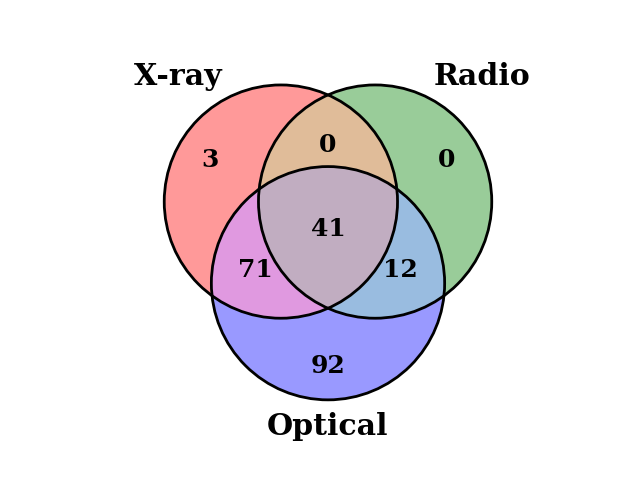}
\caption{Venn diagram of the current multi-wavelength sample of SNRs in M33. Optical detections are based on elevated [SII]/H$\alpha$ ratios and come from the catalogs of L10 and LL14. X-ray detections are from this work and the work of L10. Radio detections are taken from \citet{Gordon1999}.}
\label{fig:venn}
\end{figure}

We provide an updated Venn diagram of the current multi-wavelength detections of SNRs in M33 in Figure~\ref{fig:venn}. The prevalence of optically detected SNRs with elevated [SII]/H$\alpha$ is due primarily to the the efficacy of this diagnostic ratio in separating SNRs from other contaminants. We boost the number of X-ray detections for the previously optically detected sources owing to our large survey area and increased soft sensitivity, which is particularly adept at detecting thermal X-ray emission of extended sources. The lack of SNRs detected only in X-rays is due to the difficulty in separating SNR candidates from other sources of soft X-ray emission on the basis of X-rays alone. By selecting primarily for candidates with multiwavelength confirmation (optical and X-ray), we may be missing some young, X-ray emitting SNRs in the free expansion phase. The dearth of radio detected SNRs is affected by observational limits; most radio surveys do not furnish the requisite angular resolution and sensitivity to definitively identify SNRs without optical or X-ray follow-up. In the absence of a deep radio survey of M33, the combination of optical emission-line diagnostics and X-ray detections remains the most reliable way to identify SNRs, although such detection methods may be biased against detecting SNRs hosting pulsar wind nebulae (plerion-type SNRs), SNRs that are Balmer-dominated, and oxygen-rich SNRs.

\subsection{Spectral Fitting}\label{sec:spectralfit}

We attempt to type the progenitors for 15 of the SNRs in our sample by fitting their spectra using \texttt{XSPEC}. These 15 sources are some of the brightest X-ray emitting SNRs in M33 for which there is both {\it XMM-Newton} and {\it Chandra} data, and for which there are enough on-axis counts for detailed spectral fitting. We used both {\it XMM-Newton} and {\it Chandra} data in cases where {\it Chandra} data was available and provided $\approx$ 40\% more counts
than {\it XMM-Newton} data alone (11 SNRs). All remaining SNRs were fit using {\it XMM-Newton} data only (four). For each of these SNRs we perform a simultaneous fit to both the unbinned source and background components. While this method for fitting is more computationally expensive than one in which the background is directly subtracted from the source spectrum before fitting, it avoids the problems with the non-Poisson nature of background-subtracted data and makes optimal use of the full spectral information of the SNRs. We describe the individual source and background models and fitting method in the following sections. For more details on how the spectral extractions themselves were carried out, see the Appendix. 

\subsubsection{Background Model}\label{sec:bkgmod}

Our background spectra were
fitted with a two component model: a sky component, plus a detector component.
The blank sky background was modeled as a pair of absorbed thermal plasma components ({\tt TBabs}, N$_{\rm H}$ = 5$\times$10$^{20}$ cm$^{-2}$). The best-fit low-T component had kT= 0.16 keV, and the higher temperature component had
kT= 0.75 keV. When fitting the SNRs, all blank sky background parameters were frozen, except for an overall multiplicative constant factor
which was allowed to vary for all instruments. 

For the pn, the detector component was modeled as a broad ($\sigma$ = 0.455 keV) Gaussian at 0 keV
and a broken power-law. In addition, Gaussian components were added for the detector fluorescence lines near 1.49 keV (Al-K$\alpha$), 2.15 keV (Au-L complex), 5.4 keV (Cr-K$\alpha$), 5.9 keV (Mn-K$\alpha$),
6.4 keV (Fe-K$\alpha$), 7.47 keV (Ni-K$\alpha$), 8.04 keV (Cu-K$\alpha$), 8.62 keV (Zn-K$\alpha$), and 9.6 keV (Zn-K$\beta$).
For the MOS1 and MOS2 detectors, the model consists of a pair of broken power-laws to model the
continuum component. The fluorescence lines listed above were added, along with a line at
1.75 keV (Si-K$\alpha$). The detector plus blank sky background models were fit to spectra covering most
of each detector with point sources removed. The detector fluorescence line energies were fitted, and the line widths allowed to vary. Finally, the line energies and widths were all frozen,
and the normalizations were all tied to appropriate multiples of an overall multiplicative constant factor,
one for each detector background model. In subsequent fitting, only the multiplicative constant
was allowed to vary.

Thus, the complete background model spectra have a scaling factor for the blank sky component, and separate scaling factors for each of the MOS1, MOS2, pn, and ACIS detector models. The background spectra are fit first with this model, and the results of this fit are supplied as inputs for the background component of the subsequent total (source + background) fit. This method
thus accounts for the background while performing fits to the source spectra, as opposed to subtracting any signal from the source data before fitting, which preserves the Poisson characteristics of the data and provides more reliable estimates of the
contribution of the background.

\subsubsection{Simultaneous Fitting of {\it Chandra} and {\it XMM}}\label{sec:simulfit}

We fit both the {\it Chandra} and {\it XMM-Newton} data simultaneously for 11 SNRs. The fits were 
carried out using {\tt XSPEC} with a plane-parallel shock model ({\tt vpshock}) \citep{Borkowski2001}. All parameters in the {\tt vpshock} model were tied between all instruments. In the {\tt vpshock} model the individual abundances of O, Ne, Mg, Si and Fe are allowed to vary for SNRs with a large enough number of spectral counts with all other abundances (aside from H and He) frozen to values of 0.5.
The Galactic absorption was fixed using a 
{\tt tbabs} model to a value of 0.5 $\times$ 10$^{21}$ cm$^{-2}$ \citep{Wilms2000}, while the absorption in M33 was allowed to vary using a {\tt tbvarabs} model; the metallicity for absorption in M33 was fixed at 0.5 times solar. The spectra were fit with unbinned channels and the C statistic was used.  The absorption, shock temperatures,  individual abundances, and ionization timescales were allowed to vary in the {\tt vpshock} model. The remaining 4 SNRs did not have a significant contribution from the {\it Chandra} data to their total spectral counts, and thus were fit using {\it XMM-Newton} data only, though using the same model as above. 

Each SNR was visually inspected prior to fitting to determine the number of useful spectral counts available for fitting. This involved examining SNR images, and removing fields in which spectral extraction regions fell on chip gaps or were far off axis. We find that $\textgreater$ 300 useful counts are necessary for a reliable fit, while $\textgreater$ 1200 counts are required to fit for individual abundances in a given SNR's spectrum. For the SNRs that had between 300--1200 counts, the individual abundances were frozen when performing the fit, or, O and Fe were allowed to vary, with all other abundances tied to O. 

We provide the best-fitting spectral parameters and associated 90\% confidence intervals from all of our fits
in Table~\ref{tab:fittab}. The analysis of the spectral fitting results are discussed in the following section.
  
\subsection{Spectral Fit Parameters}\label{sec:spectralresults} 

{\renewcommand{\arraystretch}{1.7} 
\setlength\tabcolsep{4pt}
\begin{table*}
\caption{Parameters from {\tt XSPEC} fits for 15 SNRs in our sample with detailed spectral fits. Column 1: source ID number in this catalog. Column 2: source counts in the 0.35-2.0 keV band. Column 3: hydrogen column density for M33  from {\tt tbvarabs} model. Column 4: plasma temperature from the {\tt vpshock} model. Column 5: the ionization timescale from the {\tt vpshock} model. Column 6: the normalization from the {\tt vpshock} model. Columns 7-11: the elemental abundances from the {\tt vpshock} model listed with respect to solar values. All units are given in the column headers.} \label{tab:fittab}
\begin{tabular}{ccccccccccc}
\hline \hline
\multicolumn{1}{|c|}{ID} &
\multicolumn{1}{|p{1.7cm}|}{0.35-2.0 keV \centering \\ Cts} &
\multicolumn{1}{|p{1.7cm}|}{N$_{H}$ \centering \\\ 10$^{22}$ cm$^{-2}$} &
\multicolumn{1}{|p{1.7cm}|}{kT$_{e}$ \centering\\ keV} &
\multicolumn{1}{|p{1.7cm}|}{$\tau$  \centering\\ 10$^{11}$~cm$^{-3}$~s} &
\multicolumn{1}{|p{1.7cm}|}{K  \centering \\ 10$^{-4}$ cm$^{-5}$} &
\multicolumn{1}{|c|}{O} &
\multicolumn{1}{|c|}{Ne} &
\multicolumn{1}{|c|}{Mg}&
\multicolumn{1}{|c|}{Si} &
\multicolumn{1}{|c|}{Fe} \\
\hline
XMM-041 & 11878 & $0.02^{+0.06}_{-0.02}$ & $0.66^{+0.03}_{-0.02}$ & $3.48^{+0.87}_{-1.02}$ & > 0.18 & $0.51^{+0.12}_{-0.15}$ & $0.56^{+0.08}_{-0.10}$ & $0.47^{+0.08}_{-0.10}$ & $0.56^{+0.15}_{-0.12}$ & $0.45^{+0.06}_{-0.08}$ \\
XMM-073 & 4347 & $0.21^{+0.12}_{-0.11}$ & $0.69^{+0.21}_{-0.17}$ & $1.35^{+0.80}_{-0.45}$ & $0.39^{+0.56}_{-0.14}$ & $0.34^{+0.15}_{-0.13}$ & $0.33^{+0.07}_{-0.06}$ & $0.86^{+0.21}_{-0.15}$ & $0.50^{+0.25}_{-0.18}$ & $0.27^{+0.22}_{-0.06}$ \\
XMM-119 & 2665 & $0.10^{+0.01}_{-0.10}$ & $0.56^{+0.13}_{-0.08}$ & $1.56^{+1.12}_{-0.88}$ & $0.42^{+0.13}_{-0.15}$ & $0.20^{+0.04}_{-0.03}$ & $0.31^{+0.08}_{-0.05}$ & $0.25^{+0.13}_{-0.08}$ & $0.48^{+0.33}_{-0.22}$ & $0.20^{+0.13}_{-0.03}$ \\
XMM-067 & 2109 & $0.11^{+0.19}_{-0.11}$ & $0.54^{+0.05}_{-0.18}$ & $7.73^{+5.27}_{-4.31}$ & $0.34^{+0.59}_{-0.14}$ & $0.27^{+0.17}_{-0.13}$ & $0.40^{+0.19}_{-0.19}$ & $0.19^{+0.2}_{-0.15}$ & $0.48^{+0.52}_{-0.36}$ & $0.18^{+0.07}_{-0.10}$ \\
XMM-153 & 1910 & $0.10^{+0.04}_{-0.10}$ & $0.74^{+0.22}_{-0.23}$ & $0.46^{+0.85}_{-0.23}$ & $0.15^{+0.12}_{-0.05}$ & $0.26^{+0.05}_{-0.07}$ & $0.31^{+0.09}_{-0.11}$ & $0.17^{+0.17}_{-0.14}$ & $0.85^{+0.45}_{-0.50}$ & $0.34^{+0.16}_{-0.17}$ \\
XMM-068 & 1207 & $0.00^{+0.03}_{-0.00}$ & $0.48^{+0.09}_{-0.03}$ & $5.62^{+8.38}_{-1.85}$ & $0.23^{+0.04}_{-0.07}$ & $0.43^{+0.19}_{-0.07}$ & $0.57^{+0.21}_{-0.12}$ & $0.36^{+0.12}_{-0.21}$ & $0.39^{+0.51}_{-0.38}$ & $0.15^{+0.03}_{-0.03}$ \\
XMM-034 & 742 & $0.10^{+0.06}_{-0.10}$ & $0.65^{+0.89}_{-0.16}$ & $0.27^{+0.38}_{-0.12}$ & $0.03^{+0.02}_{-0.01}$ & $0.67^{+0.19}_{-0.15}$ & -- & -- & -- & $0.36^{+0.33}_{-0.21}$ \\
XMM-039 & 1820 & $0.10^{+0.13}_{-0.10}$ & $0.51^{+0.21}_{-0.18}$ & $0.9^{+2.43}_{-0.52}$ & > 0.03 & $0.39^{+0.19}_{-0.13}$ & $0.57^{+0.26}_{-0.24}$ & $0.41^{+0.50}_{-0.30}$ & $0.42^{+1.04}_{-0.42}$ & $0.29^{+0.24}_{-0.14}$ \\
XMM-082 & 2143 & <0.14 & $0.43^{+0.01}_{-0.03}$ & >50.00 & $0.14^{+0.08}_{-0.10}$ & $0.20^{+0.23}_{-0.03}$ & $0.40^{+0.13}_{-0.07}$ & $0.70^{+0.08}_{-0.18}$ & $0.36^{+0.29}_{-0.25}$ & $0.10^{+0.02}_{-0.02}$ \\
XMM-151 & 937 & $0.10^{+0.04}_{-0.10}$ & $0.36^{+0.24}_{-0.14}$ & $1.04^{+1.13}_{-0.03}$ & >0.39 & $0.48^{+0.23}_{-0.14}$ & -- & -- & -- & $0.50^{+0.32}_{-0.22}$ \\
XMM-066 & 982 & $0.26^{+0.00}_{-0.01}$ & $0.6^{+0.03}_{-0.4}$ & $0.17^{+0.07}_{-0.01}$ & >0.37 & $0.54^{+0.06}_{-0.13}$ & -- & -- & -- & $1.20^{+1.17}_{-0.64}$ \\
XMM-132 & 1070 & $0.10^{+0.04}_{-0.10}$ & $0.73^{+0.59}_{-0.28}$ & $0.25^{+0.34}_{-0.15}$ & >0.06 & $0.64^{+0.24}_{-0.11}$ & -- & -- & -- & $0.60^{+0.53}_{-0.27}$ \\
XMM-065 & 502 & $0.28^{+0.31}_{-0.19}$ & $0.92^{+1.59}_{-0.56}$ & $0.14^{+0.21}_{-0.06}$ & $0.03^{+-0.02}_{-0.01}$ & -- & -- & -- & -- & -- \\
XMM-136 & 585 & $0.10^{+0.05}_{-0.10}$ & $0.38^{+0.05}_{-0.07}$ & >6.72 & >0.09 & $0.24^{+0.07}_{-0.13}$ & -- & -- & -- & $0.15^{+0.08}_{-0.07}$ \\
XMM-118 & 310 & $0.10^{+0.14}_{-0.10}$ & $0.31^{+0.32}_{-0.07}$ & >19.78 & $0.04^{+0.03}_{-0.03}$ & $0.56^{+0.77}_{-0.36}$ & -- & -- & -- & $0.31^{+0.3}_{-0.17}$ \\
\hline
\end{tabular}
\end{table*}

The parameters from the detailed spectral fitting are recorded in Table~\ref{tab:fittab} for the model described in Section~\ref{sec:simulfit}. The fitted parameters include the hydrogen column density (N$_{H}$), the electron temperature (kT$_{e}$), the ionization timescale ($\tau$), and the normalization (K). The abundances of O, Ne, Mg, Si and Fe are reported relative to the solar value, and are allowed to vary for SNRs with $\textgreater$ 1200 counts. An example of the model fit to the data for a single SNR (XMM-041, L10-025) is plotted in Figure~\ref{fig:snrfitex}. Each set of two panels illustrates the total fit, associated components (left panel), and the background fit only (right panel) for each instrument (pn, MOS1, MOS2, and ACIS). Figure~\ref{fig:snrfitex} demonstrates the robustness of the fitting technique by illustrating the strong contribution of the background component, and in particular the instrument background, at higher energies.

Along with the fitted parameters we derive
inferred physical parameters from these values, such as pre-shock H density (n$_{o}$), ionization age (t$_{ion}$), dynamical age (t$_{dyn}$), shock velocity (v$_{s}$), initial explosion energy (E$_{o}$), and swept-up mass (M$_{su}$). These parameters are calculated based on a Sedov model, and assuming a volume-filling factor of one, spherically symmetric SNRs, strong shock jump conditions, and electron-ion equilibrium. The effects of some of these simplifying assumptions are discussed in the following paragraphs.
The values for the above calculated physical parameters are reported for each SNR in Table~\ref{tab:phystab} and represent spatially averaged quantities over the SNR. The {\tt XSPEC} model fits provide estimates for the electron temperature (kT$_{e}$), the ionization timescale ($\tau$), and the normalization (K) (see Table~\ref{tab:fittab}). 

Following \citet{Hughes1998} and \citet{Gaetz2007} we can calculate the Sedov model parameters in Table~\ref{tab:phystab} based on the observational values for the electron temperature, ionization timescale, normalization, and SNR radii. We use the radius for each SNR reported from L10 in our calculations (see Table~\ref{tab:sourcetab}), and assume errors of $\sim$ 9\% on the reported radii, based on adding in quadrature an error of 5\% in angular size and assuming azimuthal asymmetry, and an error of 7\% in the assumed distance to M33. Errors on the derived physical quantities are calculated from the 90\% confidence interval on the fitted parameters using 10,000 Monte Carlo draws from the error distribution, and the propagation of the previously stated errors on the radii and angular sizes of the SNRs.

\newpage
 \newgeometry{margin=1cm}
\begin{landscape}
\begin{figure}
    \centering
    \begin{subfigure}[pn]
        \centering
        \includegraphics[scale=0.3]{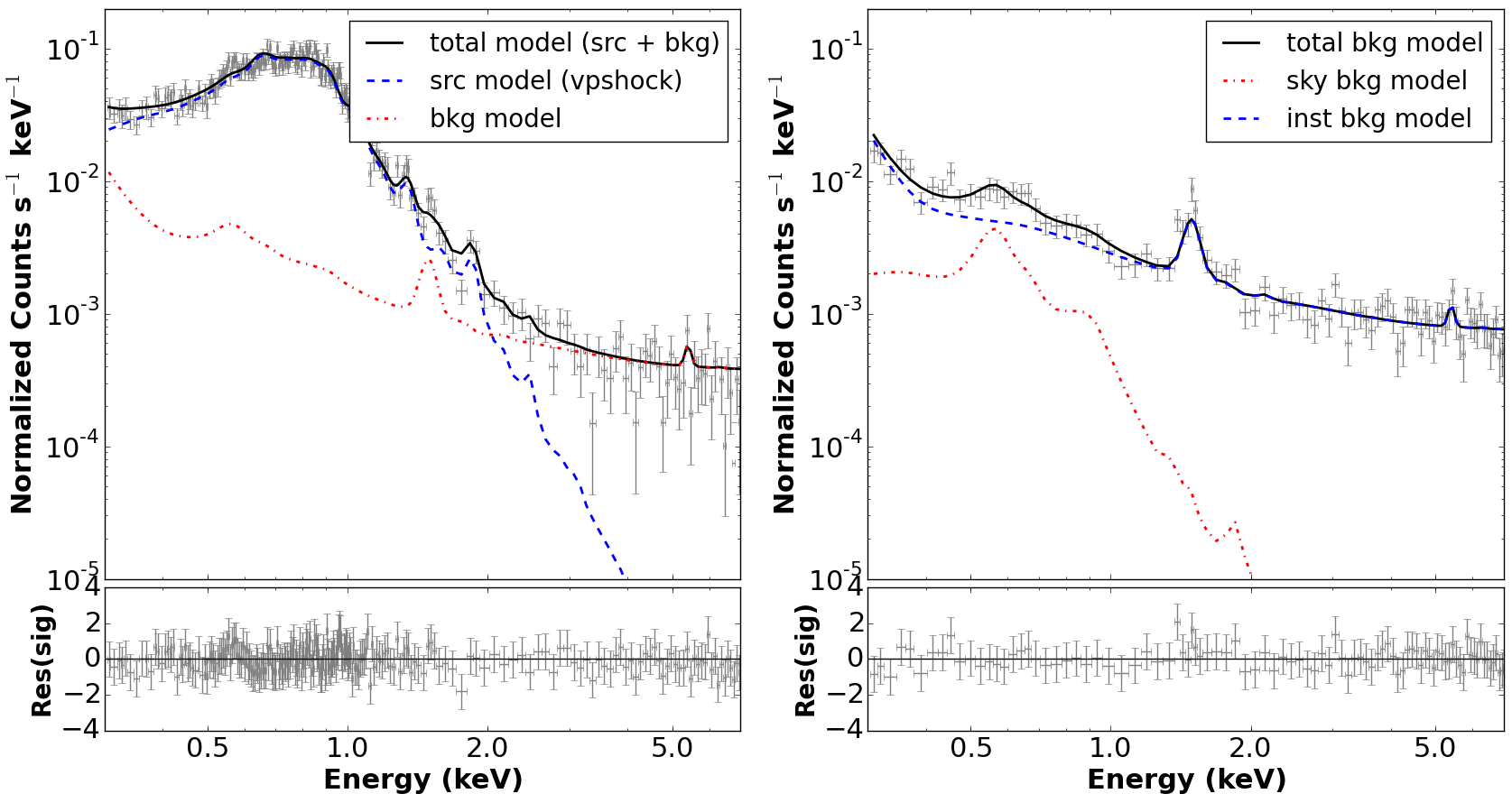}
    \end{subfigure}
    \begin{subfigure}[MOS1]
        \centering
        \includegraphics[scale=0.3]{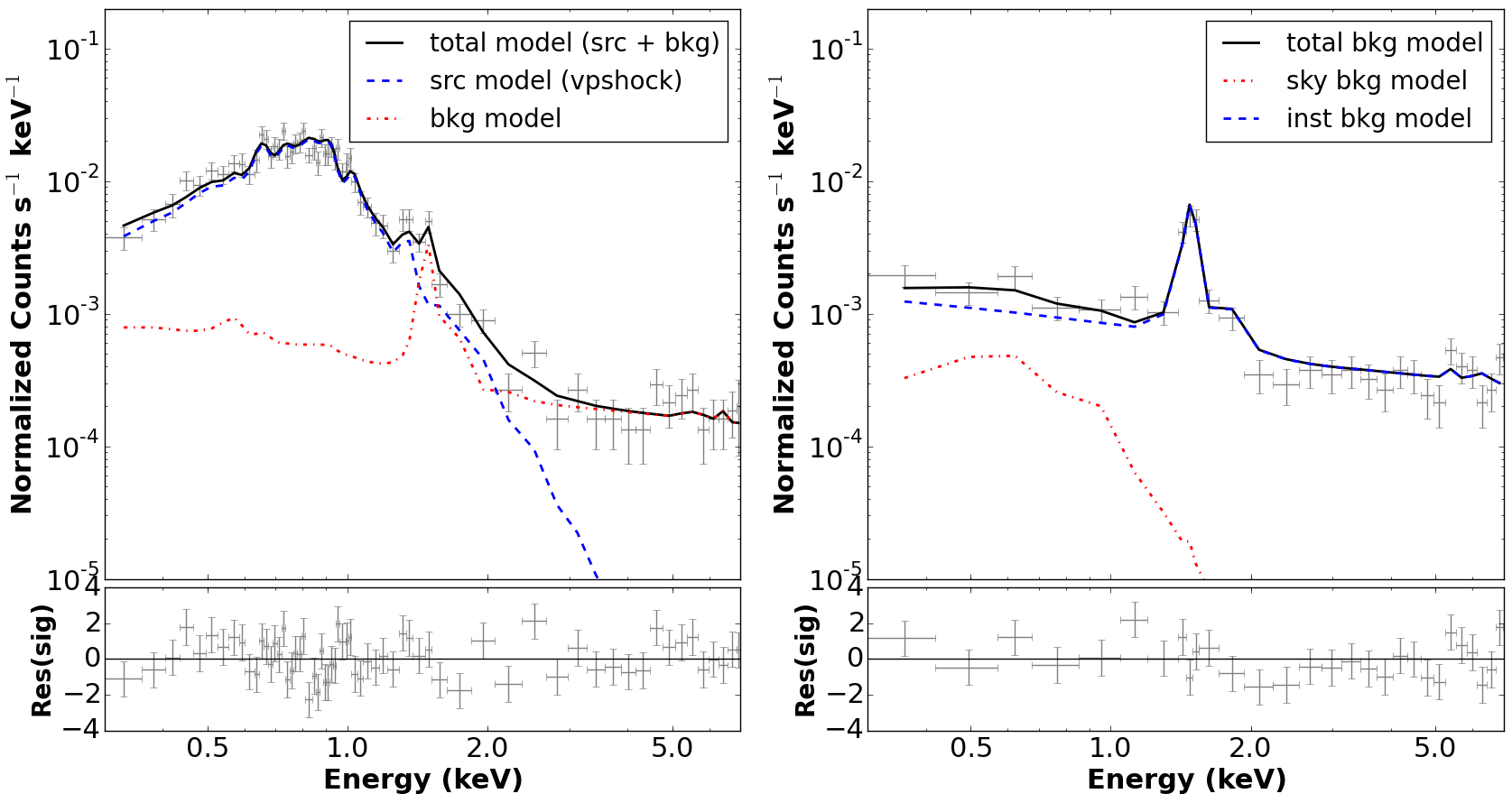}
    \end{subfigure}
     \begin{subfigure}[MOS2]
        \centering
        \includegraphics[scale=0.3]{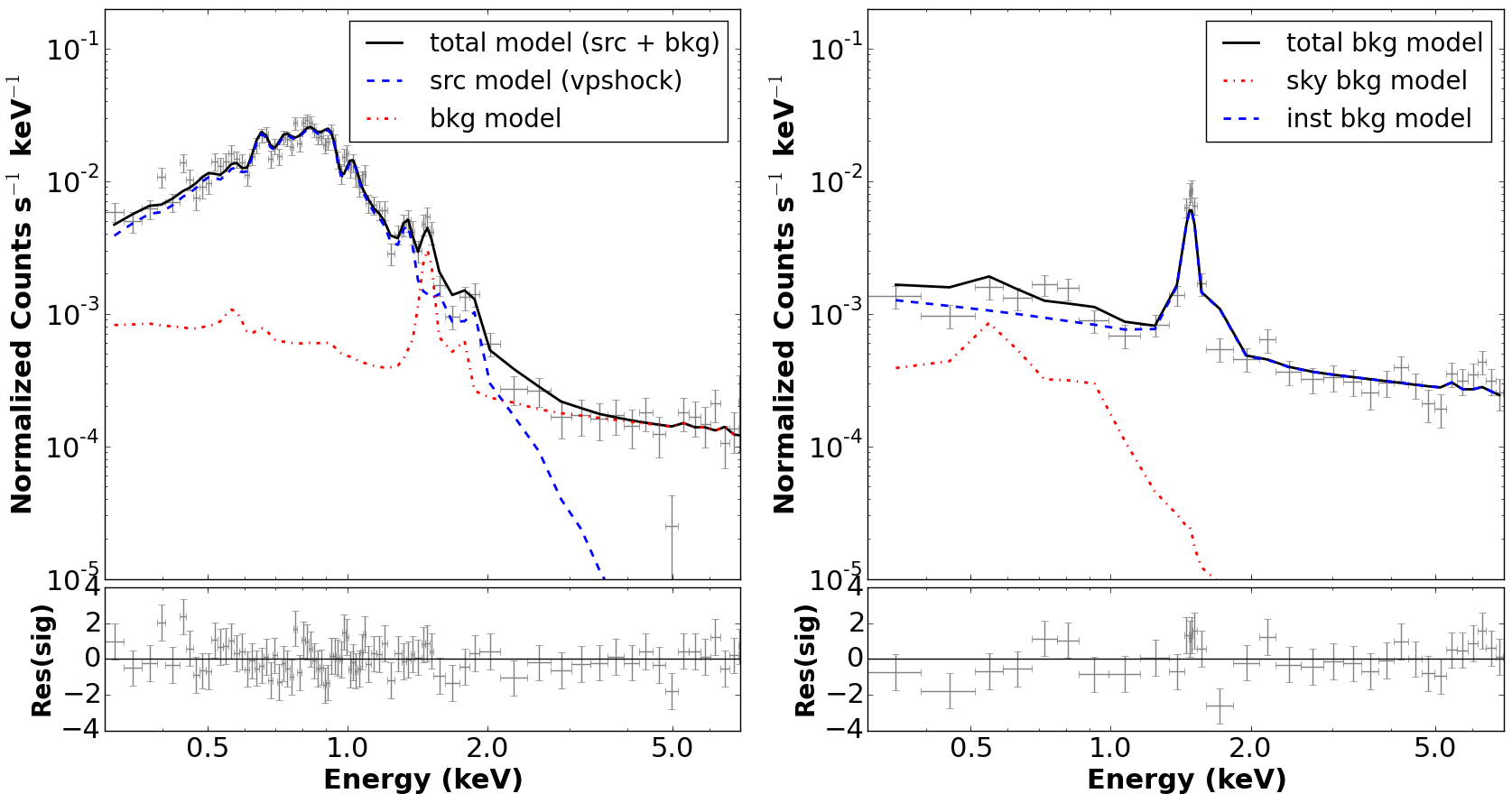}
    \end{subfigure}
     \begin{subfigure}[ACIS]
        \centering
        \includegraphics[scale=0.3]{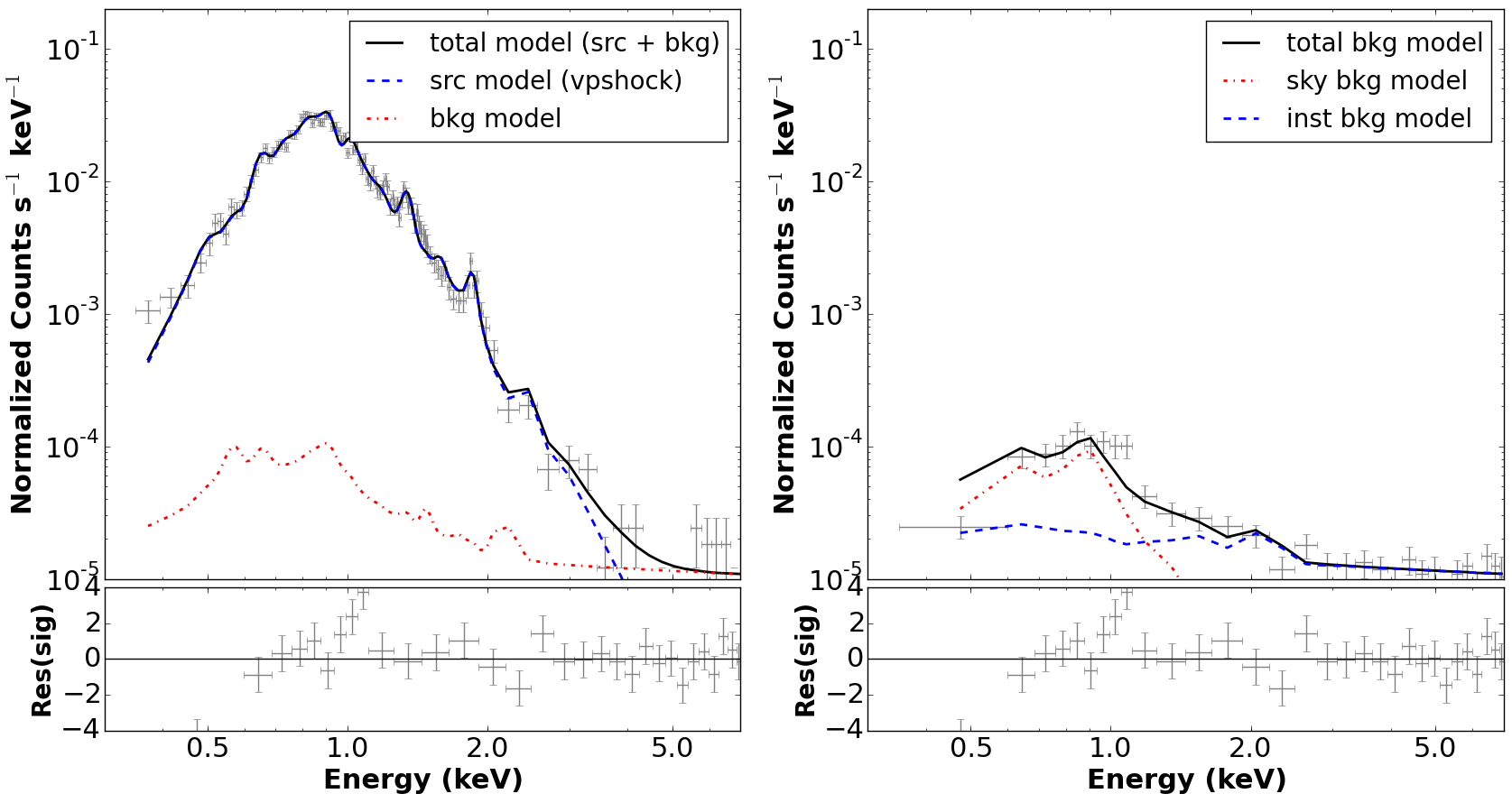}
    \end{subfigure}
    \caption{Sample fit for the SNR XMM-041 (L10-025) using a {\tt vpshock} model for the source with O, Ne, Mg, Si and Fe abundances as free parameters. Panels a, b, c, and d represent fits the individual cameras: pn, MOS1, MOS2, and ACIS, respectively. Within each panel, the left-hand panel depicts the total fitted model in black, with the total model components, source ({\tt vpshock}) and background shown in blue and red, respectively. The gray points show the data and residuals with associated errorbars. The right-hand panel shows the total fitted background model in black, and its components, sky and instrument background, in blue and red. The data and residuals are again plotted as gray points, with associated errorbars.}\label{fig:snrfitex}
\end{figure}
\end{landscape}
\restoregeometry 
\newpage

The pre-shock H density is calculated from the fitted normalization, distance to M33, and measured radii using the following equation from \citet{Gaetz2007}:

\begin{equation}
n_{o} = 1.58 K_{-4}^{1/2}D_{800}R_{s,10}^{3/2}~~cm^{-3}
\end{equation}

\noindent where $K_{-4}$ is the normalization in units of 10$^{-4}$ cm$^{-5}$, D$_{800}$ is the assumed distance in units of 800 kpc, and R$_{s,10}$ is the source radius in units of 10 pc. Assuming strong shock jump conditions, this pre-shock H density can be used to calculate the post-shock density as follows: $ n_{e} = 4.8\times n_{o}$. The post-shock density, coupled with the ionization timescale ($\tau$) gives an estimate of the ionization age. Following \citet{Borkowski2001}, we note that at a given shock velocity the ionization timescale, $\tau$, from a plane-parallel shock model, such as the {\tt vpshock} model used in our fits, may be considerably larger than the emission-averaged ionization timescale, $\textless$$\tau$$\textgreater$, in the Sedov model. We calculate the ionization ages in Table~\ref{tab:phystab} from the plane-parallel shock $\tau$, and note that the ages based on the Sedov $\textless$$\tau$$\textgreater$ may be smaller, leading the ages reported in Table~\ref{tab:phystab} to be an upper limit. The dynamical ages are proportional to the observed radii divided by the shock velocity  ($v_{s} \sim \sqrt{kT_{e}}$), and we find that they are systematically much larger than the ages calculated based on the ionization timescale from the fits. This may imply that the simplifying assumptions of all SNRs in the Sedov phase and in electron-ion equilibrium as applied to our constant-temperature plane-parallel shock model fits are not appropriate for all SNRs. Despite this, we proceed with these assumptions to provide estimates of the explosion energies and swept-up masses for each fitted SNR. We derive an average explosion energy for all SNRs of $\sim$ 1.5 $\times$ 10$^{51}$~ergs and swept-up masses that are all on the order of hundreds M$_{\odot}$. There is an uncertainty in the swept-up masses due to the distribution of circumstellar and interstellar material around the progenitor that is not accounted for here. The Sedov model assumes a point explosion in a uniform medium, however for a CC SN
the stellar winds of the progenitor will have sculpted the surrounding medium, meaning that the density currently being encountered
by the blast wave may be larger than the density encountered at an earlier stage, leading to a general overestimate of the swept-up mass. For this reason, we consider the swept-up mass calculations to be an upper bound.

\begin{figure}
\centering
\includegraphics[trim=0 20 20 60,clip, scale=0.4]{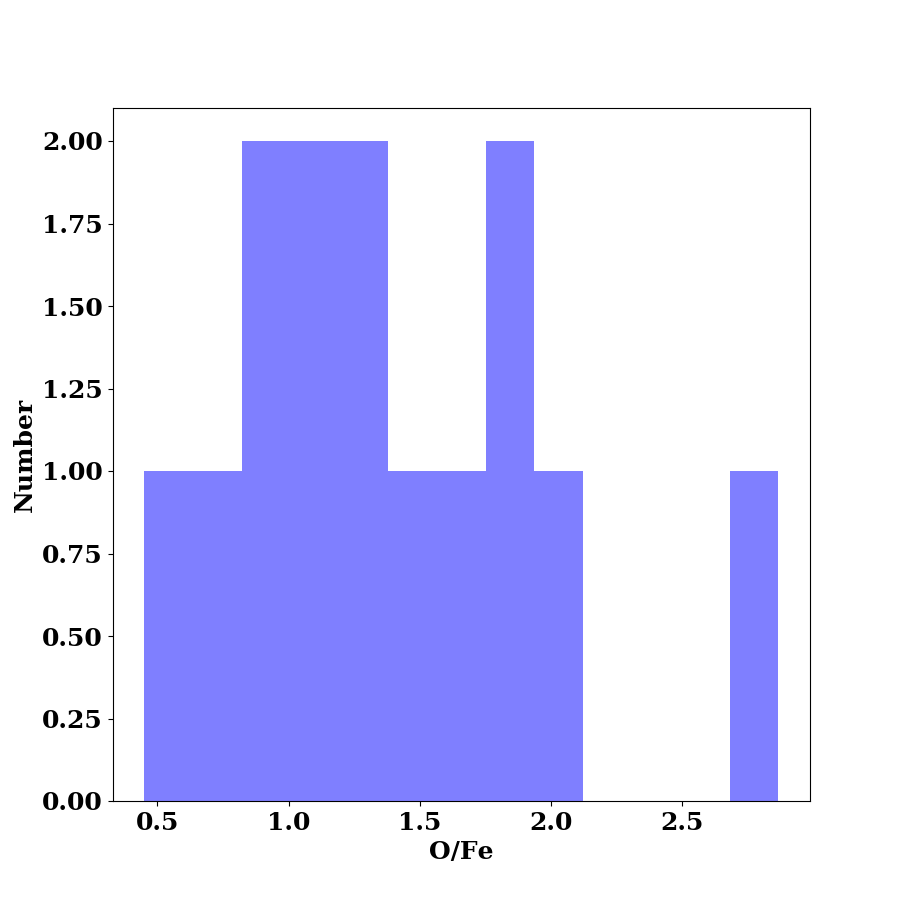}
\caption{Histogram of the O/Fe values for all 15 fitted SNRs. There is one clear outlier (XMM-068) from the overall distribution, with highly elevated O/Fe, which may be indicative of CC ejecta enrichment in the vicinity of this SNR.}
 \label{fig:OFehist}
\end{figure} 

\begin{figure}
\centering
\includegraphics[trim=80 40 70 80,clip,scale=0.27]{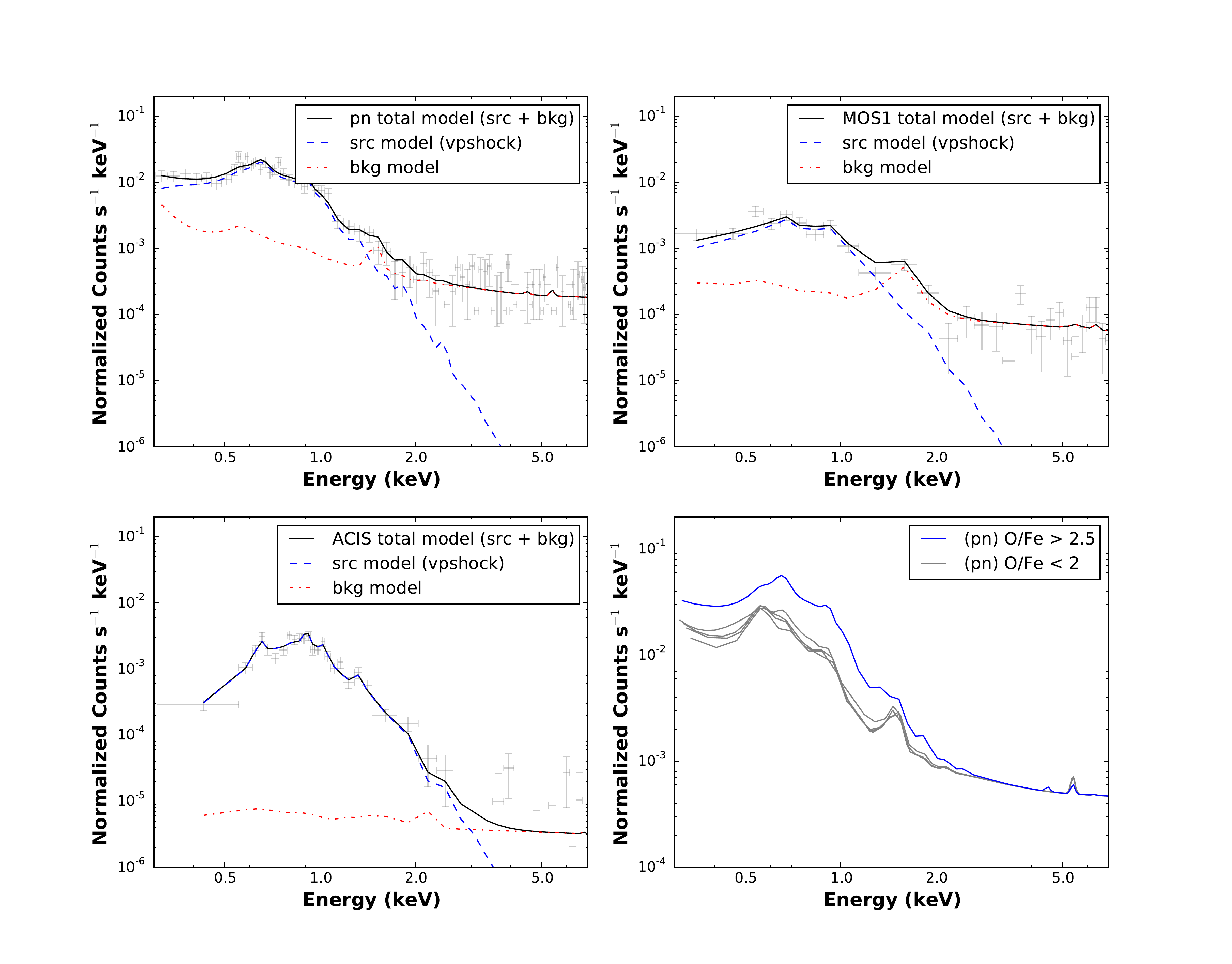}
\caption{{\tt XSPEC} {\tt vpshock} model fit for SNR XMM-068 for the pn (top-left), MOS1 (top-right), and ACIS (bottom-left) instruments. The model components are labeled in each panel. The bottom right panel compares the fitted pn spectrum for XMM-068 to the pn spectra for four SNRs in this sample with similar numbers of counts, but lower O/Fe values. All spectra have been normalized to have the same number of counts at 5~keV.}
 \label{fig:L10_037}
\end{figure}

\begin{table*}
\centering
\caption{Physical parameters for SNRs calculated from fitted parameters in Table~\ref{tab:fittab}. Column 1: source ID number in this catalog. Column 2: pre-shock H density. Column 3: ionization age for the SNR. Column 4: dynamical age for the SNR. Column 5: shock velocity. Column 6: explosion energy. Column 7: swept-up mass. All units are given in the column headers.} \label{tab:phystab}
\begin{tabular}{ccccccc}
\hline \hline
\multicolumn{1}{|c|}{ID} &
 \multicolumn{1}{|p{1.7cm}|}{n$_{o}$ \centering \\ cm$^{-3}$} &
  \multicolumn{1}{|p{1.7cm}|}{ t$_{ion}$\centering \ 10$^{3}$ yrs} &
 \multicolumn{1}{|p{1.7cm}|}{ t$_{dyn}$\centering \\ 10$^{3}$ yrs} &
 \multicolumn{1}{|p{1.7cm}|}{ v$_{s}$\centering \\ km~s$^{-1}$} &
  \multicolumn{1}{|p{1.7cm}|}{ E$_{o}$ \centering \\ 10$^{51}$ erg} &
   \multicolumn{1}{|p{1.7cm}|}{ M$_{su}$ \centering \\ M$_{\odot}$} \\
\hline
XMM-041& >0.40 & $5.00^{+1.20}_{-5.00}$ & $6.70^{+0.40}_{-0.50}$ & $740^{+20}_{-20}$ & >1.50 & >180 \\
XMM-073 & $2.39^{+1.44}_{-0.57}$ & $0.40^{+0.30}_{-0.20}$ & $3.40^{+0.60}_{-0.50}$ & $760^{+110}_{-100}$ & $1.00^{+0.80}_{-0.40}$ & $120^{+80}_{-40}$ \\
XMM-119 & $1.04^{+0.24}_{-0.25}$ & $1.00^{+0.90}_{-0.50}$ & $5.80^{+0.60}_{-0.70}$ & $680^{+80}_{-50}$ & $1.60^{+0.60}_{-0.50}$ & $230^{+70}_{-80}$ \\
XMM-067  & $1.17^{+0.81}_{-0.30}$ & $4.40^{+3.70}_{-2.70}$ & $5.30^{+1.30}_{-0.40}$ & $670^{+30}_{-130}$ & $1.20^{+0.80}_{-0.50}$ & $180^{+140}_{-60}$ \\
XMM-153 & $0.78^{+0.30}_{-0.17}$ & $0.40^{+0.70}_{-0.20}$ & $4.60^{+1.00}_{-0.60}$ & $780^{+110}_{-140}$ & $1.10^{+0.60}_{-0.40}$ & $120^{+60}_{-40}$ \\
XMM-068& $0.30^{+0.06}_{-0.07}$ & $12.30^{+20.50}_{-4.10}$ & $10.00^{+0.70}_{-1.00}$ & $630^{+60}_{-30}$ & $2.10^{+0.70}_{-0.60}$ & $350^{+100}_{-110}$ \\
XMM-034 & $0.16^{+0.06}_{-0.04}$ & $1.10^{+1.80}_{-0.50}$ & $7.30^{+1.20}_{-2.60}$ & $730^{+400}_{-100}$ & $0.80^{+1.20}_{-0.30}$ & $110^{+50}_{-40}$ \\
XMM-039  & >0.15 & $2.70^{+6.50}_{-2.70}$ & $7.60^{+1.90}_{-1.30}$ & $650^{+130}_{-130}$ & >0.40 & >70 \\
XMM-082& $0.27^{+0.09}_{-0.11}$ & >103.40 & $9.90^{+0.80}_{-0.60}$ & $600^{+10}_{-30}$ & $1.30^{+0.50}_{-0.60}$ & $250^{+100}_{-110}$ \\
XMM-151  & >1.36 & $0.30^{+0.5}_{-0.3}$ & $7.2^{+2.0}_{-1.7}$ & $550^{+160}_{-120}$ & >1.1 & >270 \\
XMM-066 & >1.02 & $0.10^{+0.00}_{-0.10}$ & $5.30^{+3.00}_{-0.30}$ & $710^{+20}_{-260}$ & >0.90 & >180 \\
XMM-132 & >0.14 & $0.90^{+1.20}_{-0.90}$ & $8.70^{+2.50}_{-2.30}$ & $780^{+270}_{-170}$ & >1.40 & >180 \\
XMM-065& $0.11^{+0.03}_{-0.03}$ & $0.90^{+1.50}_{-0.40}$ & $7.30^{+3.90}_{-2.90}$ & $870^{+570}_{-310}$ & $1.40^{+2.60}_{-0.90}$ & $130^{+50}_{-50}$ \\
XMM-136& >0.18 & >31.30 & $11.30^{+1.40}_{-1.00}$ & $560^{+40}_{-60}$ & >1.00 & >200 \\
XMM-118 & $0.29^{+0.11}_{-0.12}$ & >46.20 & $8.20^{+1.30}_{-2.50}$ & $510^{+220}_{-70}$ & $0.30^{+0.40}_{-0.10}$ & $80^{+40}_{-40}$ \\
\hline
\end{tabular}
\end{table*}

The large swept-up masses for the fitted sample imply that the majority of these SNRs are older and therefore ISM-dominated. Given the derived ages and swept-up masses, we expect the SNR ejecta to be well-mixed with the surroundings, leading to fitted abundances that more closely resemble that of their surroundings, as opposed to the ejecta distributions expected for individual Type Ia or CC SNe. Because of this, we are unable to definitively type any of the SNRs in our fitted sample. We do note, however, that one of the SNRs in this fitted sample---XMM-068 (L10-037, LL14-062)---has an elevated O/Fe value of $\sim$ 3, which is markedly in excess of the O/Fe ratios for the rest of the fitted sample as demonstrated in Figure~\ref{fig:OFehist}. In a CC SNR, we would expect that if O is enhanced, Ne and Mg would also be enhanced. Because Ne line energies overlap Fe L-shell lines there will be some blending at the resolution of the CCD, and one might expect an elevated O/Fe value simply based on an anti-correlation between Ne and Fe. We do not see a strong anti-correlation between the fitted Ne and Fe values for this SNR based on a contour plot generated with the {\tt XSPEC} {\tt steppar} command, and further the Ne value is not overly abundant in the fit compared to the expected M33 metallicity abundance. This implies that the Fe abundance is actually deficient compared to O in this SNR, as opposed to simply {\it appearing} less abundant due to being masked by an enhanced Ne abundance based on lines at energies similar to those of the Fe L-shell lines which are not resolved at CCD resolution. We plot the spectral fit results for this SNR in Figure~\ref{fig:L10_037} for the pn, MOS1, and ACIS instruments, as well as for the pn compared to four other pn spectra for SNRs with O/Fe $\textless$ 2 (bottom-right panel). Qualitatively, there is no marked difference between the high O/Fe SNR (blue line), and the rest of the sample (grey lines), though without more SNRs at higher O/Fe values it is difficult to classify based on spectral shape alone. 

Based on the swept-up mass of SNR XMM-068, which is in excess of 300~M$_{\odot}$, we note that it is most likely ISM-dominated, but may still originate from an environment with generally more CC ejecta enrichment. Interestingly, XMM-068 is the only SNR in the fitted sample that was given a progenitor classification of Type Ia by LL14 based on the surrounding stellar population. By contrast, we see evidence that O is enriched compared to Fe, even within the errors, for XMM-068, suggesting that this SNR's environment contains relics of more high-mass, CC ejecta. 

The only SNR to be designated as CC in L10 is XMM-073 (L10-039, LL14-067) based on elevated  O, Ne, and Mg as compared to Fe from an X-ray spectral fit. We see only slight enhancement of O, Ne, and Mg relative to Fe in our fits for XMM-073, but L10's classification is consistent with our spectral fits within the errors on the fitted abundances. However, the light element abundance enhancements relative to iron coupled with the large swept-up mass preclude the possibility of assigning a progenitor type based on our spectral fits alone. 

M33, a star-forming Scd galaxy, is expected to have a much higher fraction of CC SNe than Type Ias. \citet{Mannucci2005} report a Type Ia SNe rate of 0.17$^{+0.068}_{-0.063}$ per century 
per 10$^{10}$ M$_{\odot}$ and a CC SNe rate of 0.86$^{+0.319}_{-0.306}$ per century per 10$^{10}$ M$_{\odot}$. Given these rates, we expect about 17\% of the SNRs in M33 to be Type Ias, or around 37 of the 218  candidates. Therefore, we expect the vast majority of our 
sample to be CC SNe, and in particular we would expect only $\sim$ 2 of the sample of bright SNRs with detailed spectral fits to be of Type Ia. LL14 tentatively type XMM-068 with this designation based on the surrounding stellar population, but we find no evidence in the X-ray spectral fits to support this designations (e.g. broad Fe L-shell complexes), and in fact we find evidence of enhanced O/Fe, though this SNR is likely ISM-dominated. XMM-066 (L10-035) is the only fitted SNR for which we measure O/Fe $\textless$ 0.5, but this SNR was left out of the LL14 catalog due to its low [SII]/H$\alpha$ value, so it has no assigned progenitor type in that catalog, and in addition, it has an extremely high swept-up mass, implying again that it would likely not retain much of the progenitor ejecta signature.  

Finally, we compared both the fitted and derived parameters to physical quantities of the SNRs such as size, luminosity, and HR measure. In doing so, we find some evidence for a correlation between the pre-shock H density and the X-ray luminosity, as can be seen in Figure~\ref{fig:nolx}. Variations in the density of the ISM surrounding the SNR progenitor will lead to differences in the resultant SNR X-ray luminosity, with higher luminosities expected from SNRs whose progenitors explode in denser environments, so evidence of such a relation is not unexpected assuming an ISM that is not spatially uniform. One might expect a correlation between X-ray HR, SNR temperature, and SNR size, as those SNRs that are cool, and evolved will have stronger emission in the soft band and be larger in size, while hotter SNRs will have more emission in the medium band. Only very high temperature, young SNRs will have strong hard band emission \citep{Maggi2016}.

Ultimately, we do not find any significant correlations between HR1 (M-S/(H+M+S)) and the temperatures, nor between the fitted temperatures and SNR sizes, as illustrated in Figure~\ref{fig:hrtemp}, though the errors on the fitted temperatures are similar to the range of measured values. Nearly all of the fitted SNRs have abundances that are similar to the M33 ISM abundance. This implies that most of the bright SNRs are ISM-dominated, and no longer display strong evidence of the ejecta signature of their progenitor. This may also indicate that fitting a global model to the entire SNR---as opposed to separately fitting and analyzing individual features---will yield fitted parameters that represent global averages of the ejecta structure of the SNR, which generally leads to some loss of information.  

\begin{figure}
\centering
\includegraphics[trim=0 20 0 80,clip, scale=0.37]{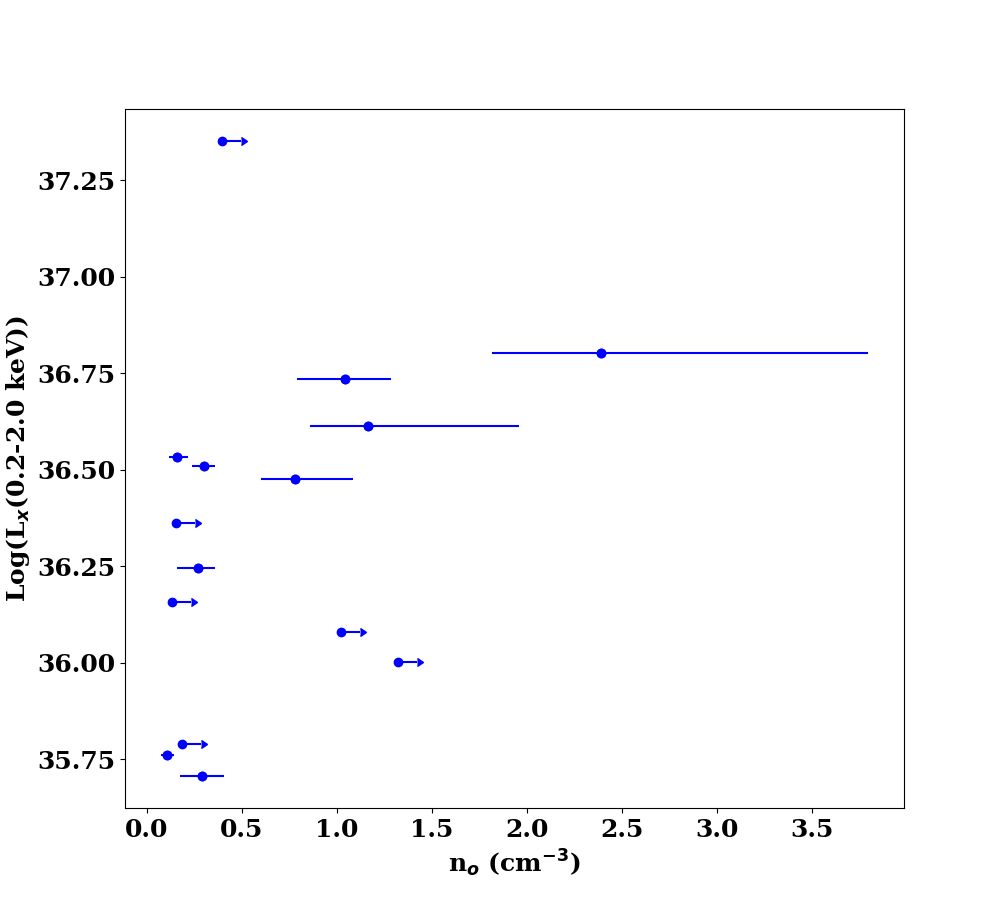}
\caption{Pre-shock H densities derived for the 15 SNRs with spectral fits versus their X-ray luminosities in the 0.2-2.0~keV  band.}
 \label{fig:nolx}
\end{figure}

\begin{figure}
\centering
\includegraphics[trim=20 20 0 70,clip,scale=0.42]{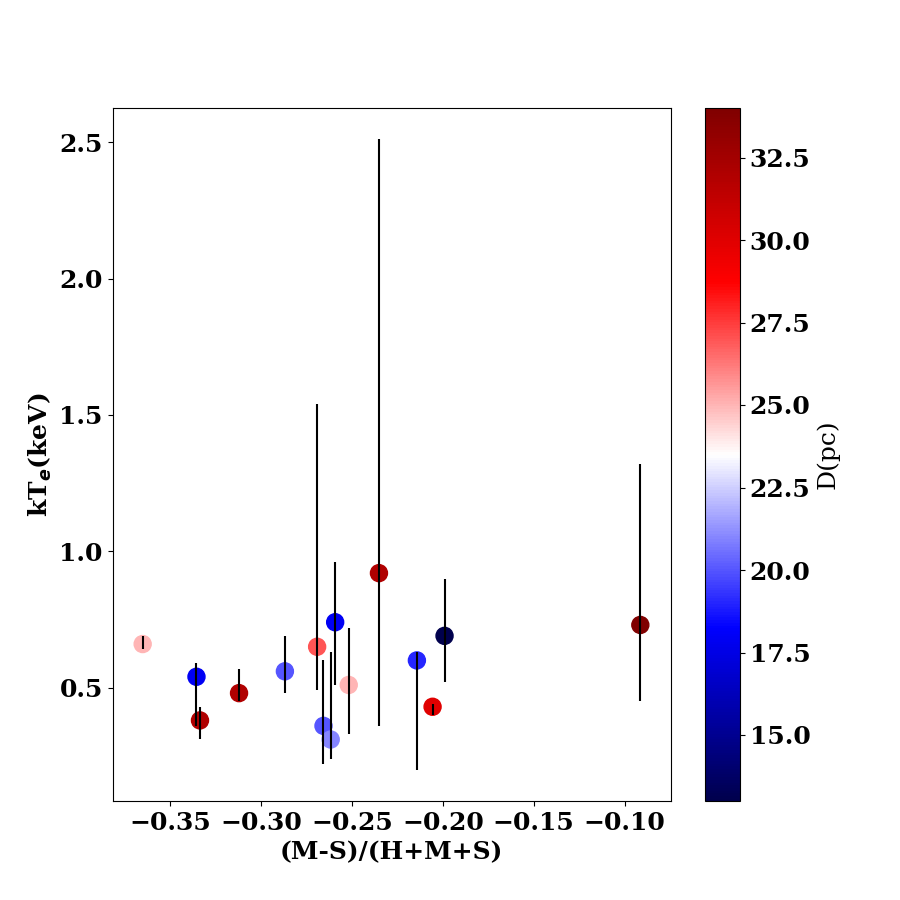}
\caption{HR1 = (M-S)/(H+M+S), where H=  1.1-4.2~keV, M = 0.7-1.1~keV, and S = 0.3-0.7~keV versus the electron temperature from spectral fits. The points are color-coded by SNR size. We find no strong correlation between SNR temperature and HR, or SNR temperature and size.}
 \label{fig:hrtemp}
\end{figure}

\subsection{Hardness Ratios}\label{sec:hr}

Hardness ratios are often used to discriminate between X-ray source types. By comparing 
X-ray fluxes across different bands one may hope to isolate different spectral shapes for sources with too few counts for reliable spectral fitting.
We attempted to type the sample of X-ray detected SNRs by HRs in bands defined in \citet{Maggi2014}. 
These bands are selected to highlight specific features in a SNR's thermal spectrum. 
The soft band is from 0.3~keV to 0.7~keV and is dominated by oxygen lines. The medium 
band ranges from 0.7~keV to 1.1~keV and includes both Fe L-shell lines indicative of a Type Ia
progenitor as well as He$\alpha$ lines from Ne~\textrm{XI} and Ne~\textrm{X}, themselves indicative of a CC progenitor. The hard band goes from
1.1--4.2~keV and is comprised of thermal continuum plus lines from Mg, Si, S, Ca, and Ar. The HRs also yield valuable information in the form of temperature; hotter SNRs should
exhibit harder HRs, while more evolved objects with cooler plasmas should be more evident in the soft band \citep{Maggi2014}. 
We calculate HRs based on counts in the soft, medium and hard bands above with the following equations: 

\begin{equation}
\centering
	HR1 = \frac{M-S}{S+M+H},~~HR2= \frac{H-M}{S+M+H}
\end{equation}

We first simulate SNR spectra in {\tt XSPEC} using a {\tt vpshock} model, a fixed Galactic absorption component ({\tt tbabs}, 0.5 $\times$ 10$^{21}$ cm$^{-2}$), and a varying M33 absorption component ({\tt tbvarabs}). We then compute the HRs from counts in the bands defined above from the simulated SNR spectra allowing only the temperature, M33 absorption component, and abundances of O and Fe to vary, with the abundances of Ne and Mg tied to the O abundance. 

\begin{figure*}
    \centering
    \begin{subfigure}
        \centering
        \includegraphics[trim=0 20 0 0,clip,scale=0.33]{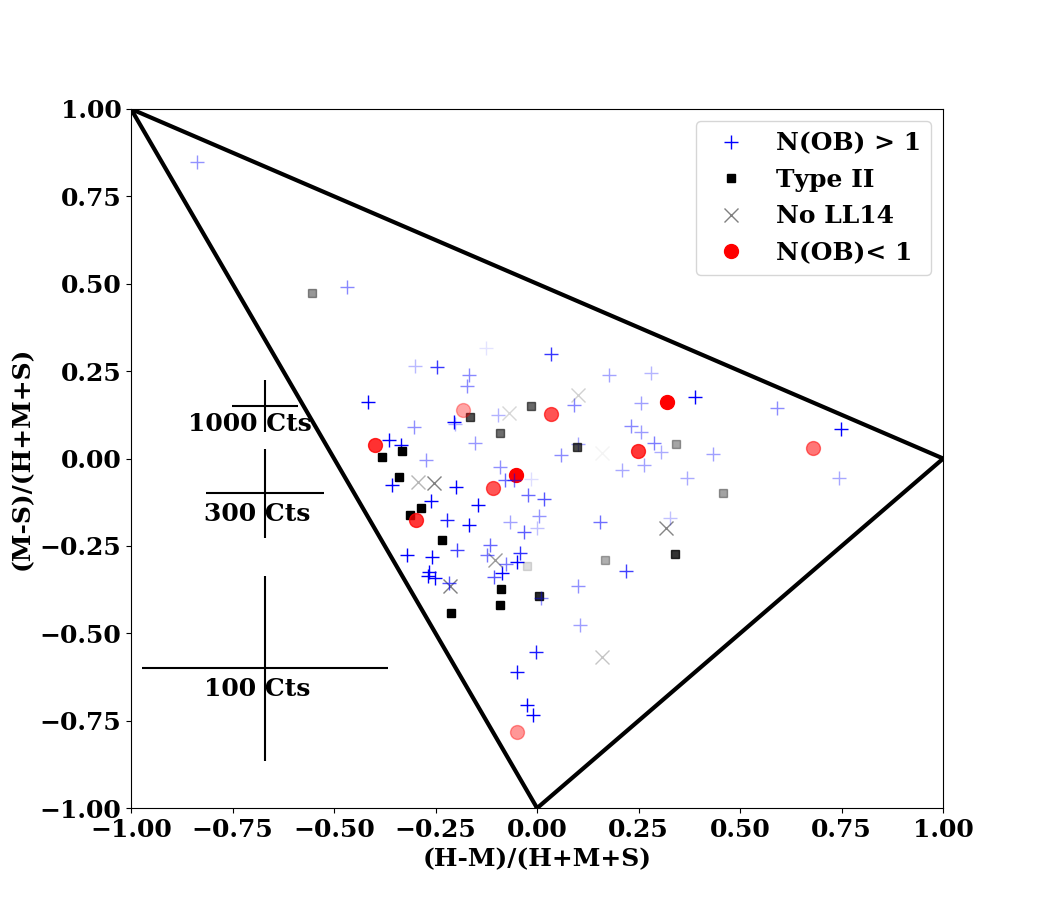}
    \end{subfigure}
    \begin{subfigure}
        \centering
        \includegraphics[trim=0 20 0 70,clip,scale=0.34]{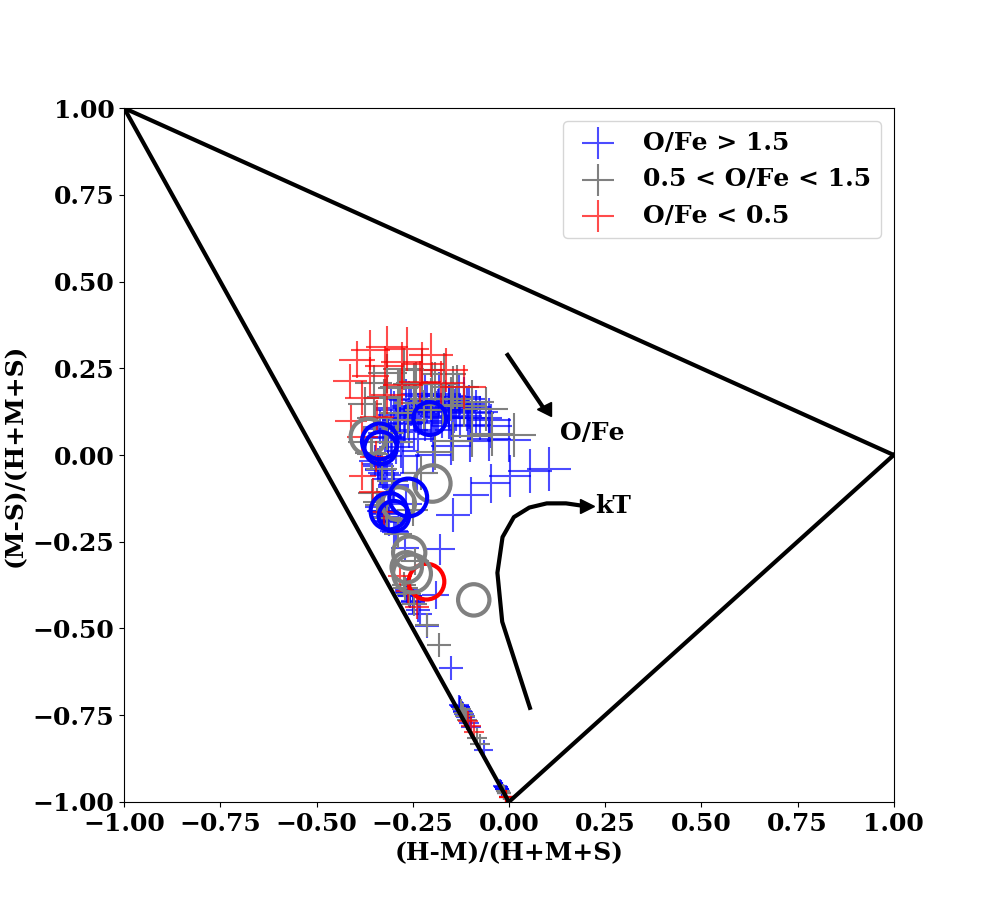}
    \end{subfigure}
          \begin{subfigure}
        \centering
        \includegraphics[trim=0 20 0 70,clip,scale=0.35]{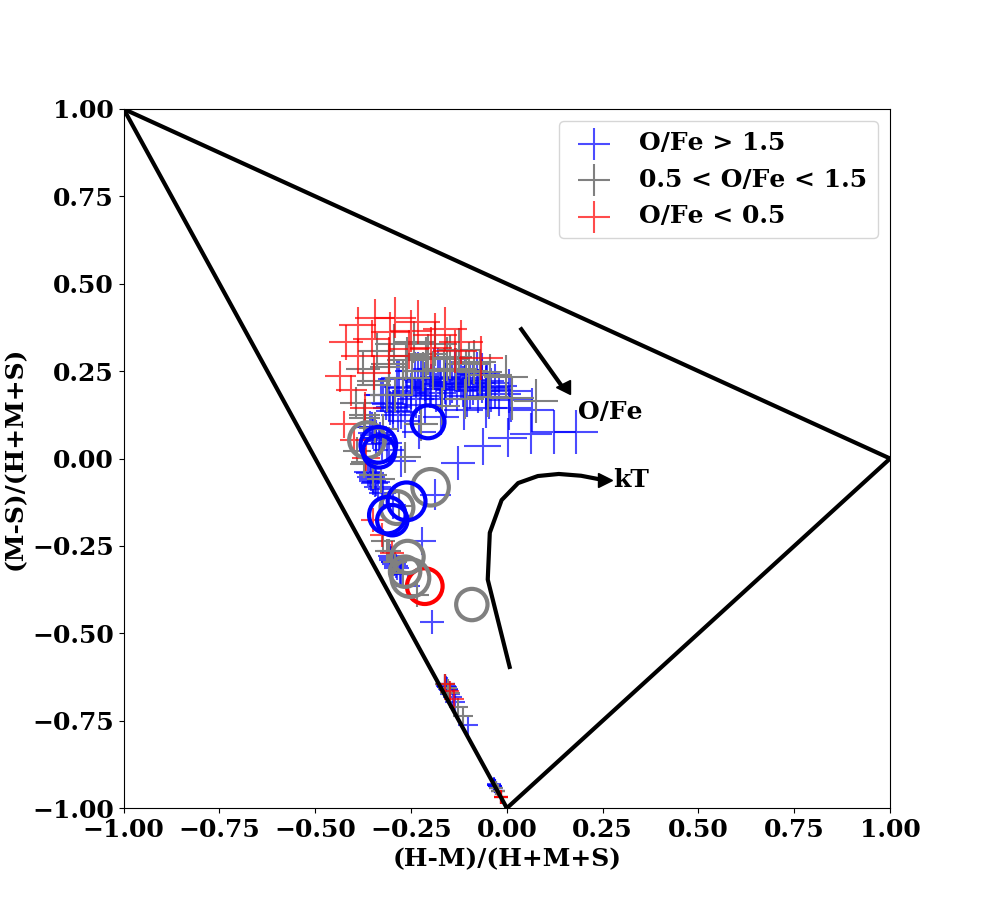}
    \end{subfigure} 
        \begin{subfigure}
        \centering
        \includegraphics[trim=0 20 0 70,clip,scale=0.35]{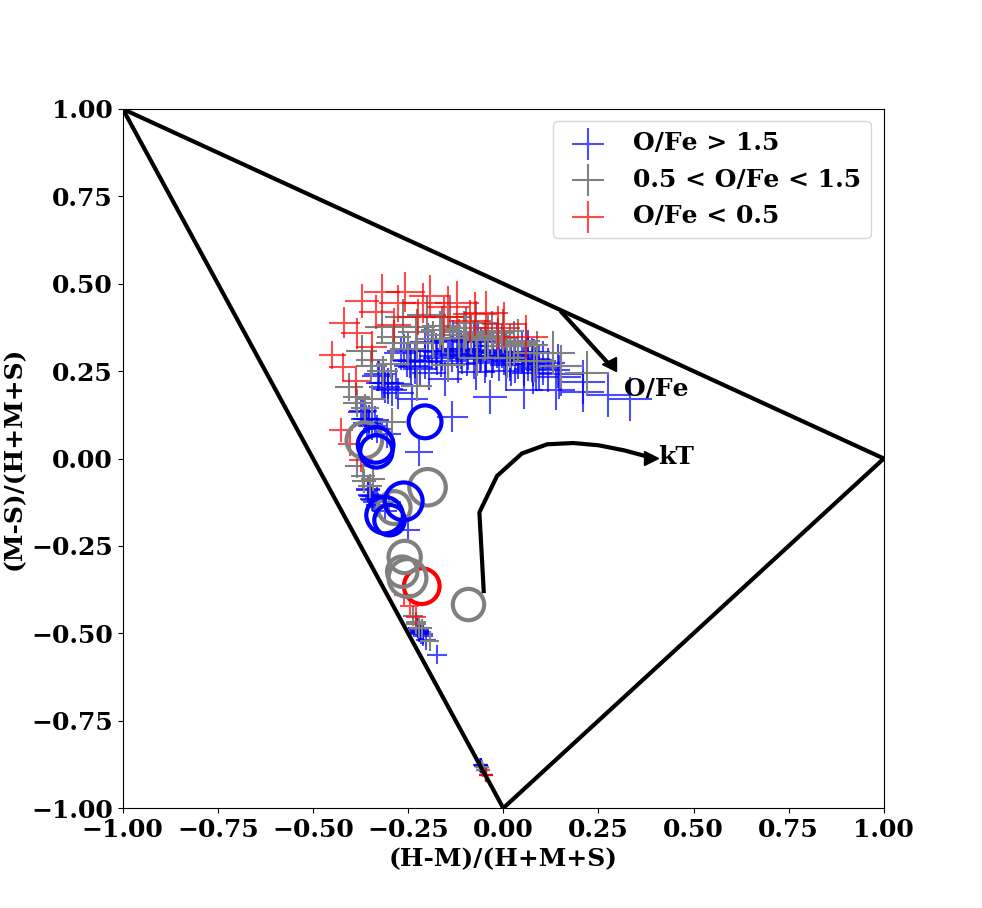}
    \end{subfigure}
\caption{{\it Top-Left}: HR from counts in the 0.3-0.7 (soft), 0.7-1.1 (medium) and 1.1-4.2 (hard)~keV bands for all SNRs detected at 3$\sigma$. Representative errors for bins of 1000 counts, 300 counts, and 100 counts are displayed for reference. Candidate Type Ia SNRs based on LL14 classifications are denoted by red circles, candidate CC SNRs from LL14 are denoted by blue crosses, and black squares are SNRs with CC progenitors based on analysis of the surrounding stellar population by \citet{Jennings2014}. Those without an LL14 or \citet{Jennings2014} match are in gray. The transparency of the points is related to signal-to-noise, with the boldest points have the highest signal-to-noise values. {\it Top-Right}: HRs from a suite of SNR spectra simulated in {\tt XSPEC} with a {\tt vpshock} model and temperatures ranging from 0.1-1.0 keV, a range of O and Fe abundances, and with a fixed M33 absorption value of N$_{H}$ = 1 $\times$ 10$^{20}$ cm$^{-2}$ (low N$_{H}$). Points are color-coded based on abundance ratio: red for low O/Fe, grey for O/Fe close to unity, and blue for elevated O/Fe. The 15 fitted SNRs are overplotted as unfilled circles using the same color scheme. The point size denotes temperature, with smaller crosses having lower temperatures. Arrows are added for reference to show the direction of increasing temperature, and increasing O/Fe ratio. {\it Bottom-Left:}:  HRs from a suite of SNR spectra simulated in {\tt XSPEC} with a {\tt vpshock} model and temperatures ranging from 0.1-1.0 keV, a range of O and Fe abundances, and with a fixed M33 absorption value of N$_{H}$ = 1.2 $\times$ 10$^{21}$ cm$^{-2}$ (intermediate N$_{H}$).  {\it Bottom-Right:}:  HRs from a suite of SNR spectra simulated in {\tt XSPEC} with a {\tt vpshock} model and temperatures ranging from 0.1-1.0 keV, a range of O and Fe abundances, and with a fixed M33 absorption value of N$_{H}$ = 3.5 $\times$ 10$^{21}$ cm$^{-2}$ (high N$_{H}$). The fitted SNRs align most closely with the simulated sample at low N$_{H}$.}
\label{fig:triangleHR_maggibands}
\end{figure*}

\begin{figure*}
    \centering
    \begin{subfigure}
        \centering
        \includegraphics[trim=0 20 0 0,clip,scale=0.33]{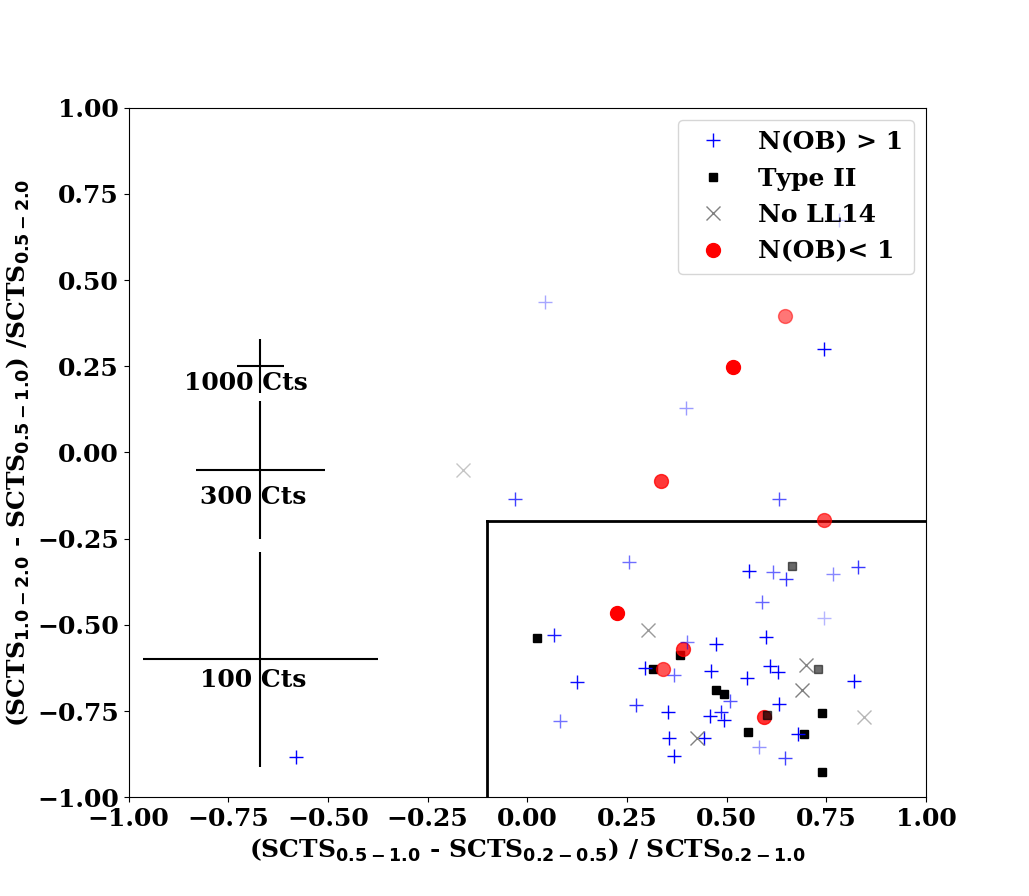}
    \end{subfigure}
    \begin{subfigure}
        \centering
        \includegraphics[trim=0 20 0 70,clip,scale=0.32]{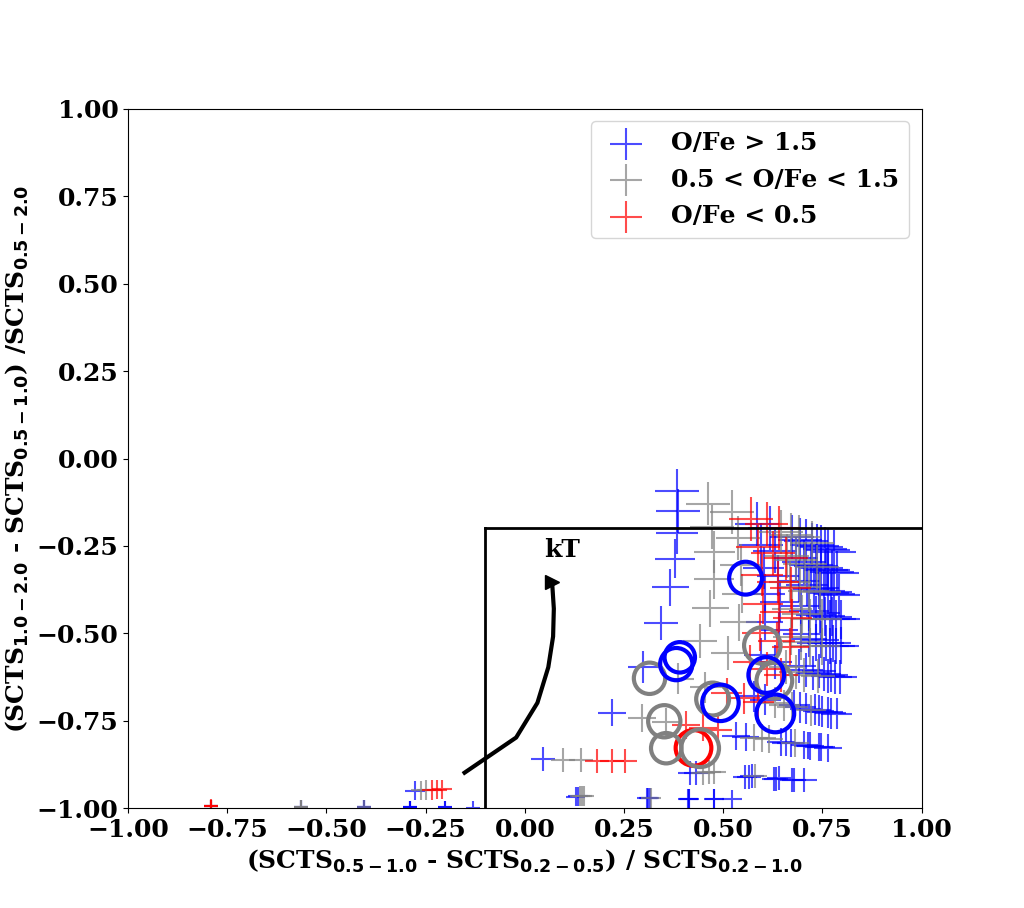}
    \end{subfigure}
      \begin{subfigure}
        \centering
        \includegraphics[trim=0 20 0 70,clip,scale=0.34]{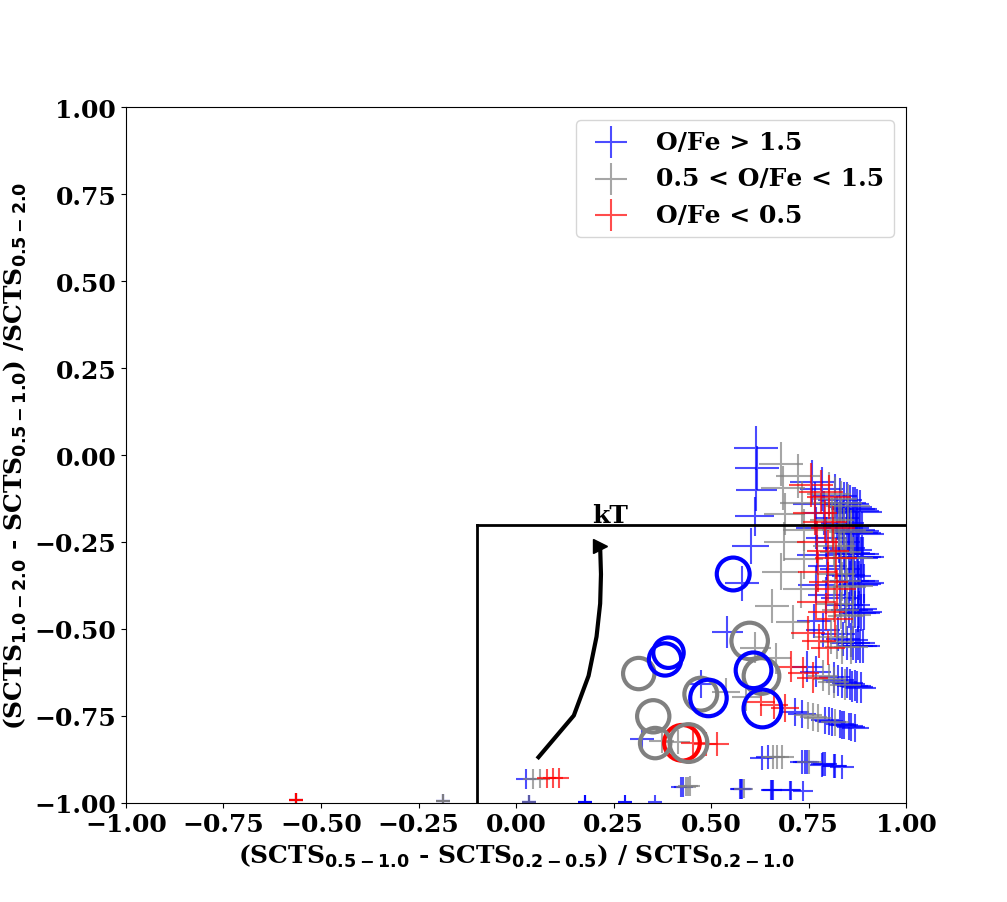}
    \end{subfigure} 
        \begin{subfigure}
        \centering
        \includegraphics[trim=0 20 0 70,clip,scale=0.33]{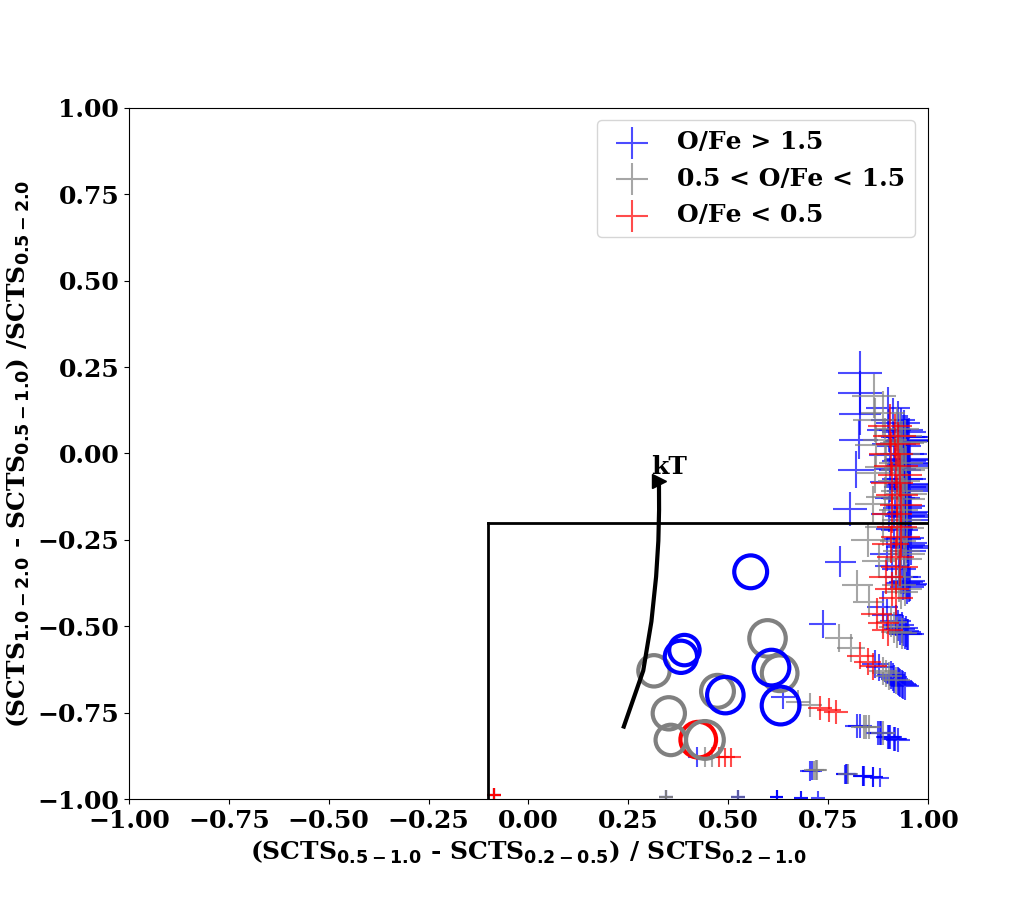}
    \end{subfigure}
\caption{{\it Top-Left}: HR from source counts for the 0.2-0.5 (soft), 0.5-1.0 (medium) and 1.0-2.0 (hard)~keV bands for SNRs detected at 3$\sigma$. Representative errors for bins of 1000 counts, 300 counts, and 100 counts are displayed for reference. Candidate Type Ia SNRs based on \citet{LeeLeeM33} classifications are denoted by red circles, while the candidate CC SNRs from this same study are denoted by blue crosses. Those without a \citet{LeeLeeM33} match are in gray. The majority of sources lie within the box defined by \citet{Pietsch2004} where we expect most SNRs to fall, but there is not clear
separation within this between Type Ia and CC SNRs. {\it Top-Right}: HRs from a suite of SNR spectra simulated in {\tt XSPEC} with a {\tt vpshock} model and temperatures ranging from 0.1-1.0 keV, a range of O and Fe abundances, and with a fixed M33 absorption value of N$_{H}$ = 1 $\times$ 10$^{20}$ cm$^{-2}$ (low N$_{H}$). SNRs are color-coded based on abundance ratio: red for low O/Fe, grey for O/Fe close to unity, and blue for elevated O/Fe. The 15 fitted SNRs are overplotted as unfilled circles using the same color scheme. The point size denotes temperature, with smaller crosses having lower temperatures. Arrows are added for reference to show the direction of increasing temperature. {\it Bottom-Left:}  HRs from a suite of SNR spectra simulated in {\tt XSPEC} with a {\tt vpshock} model and temperatures ranging from 0.1-1.0 keV, a range of O and Fe abundances, and with a fixed M33 absorption value of N$_{H}$ = 1.2 $\times$ 10$^{21}$ cm$^{-2}$ (intermediate N$_{H}$).  {\it Bottom-Right:}  HRs from a suite of SNR spectra simulated in {\tt XSPEC} with a {\tt vpshock} model and temperatures ranging from 0.1-1.0 keV, a range of O and Fe abundances, and with a fixed M33 absorption value of N$_{H}$ = 3.5 $\times$ 10$^{21}$ cm$^{-2}$ (high N$_{H}$). The fitted SNRs align most closely with the simulated sample at low N$_{H}$, though there is no clear trend with abundance ratio and HRs in these bands.}
\label{fig:boxHR_xmmbands}
\end{figure*}

The HRs computed from these simulated spectra are plotted in the top-right, bottom-left, and bottom-right panels of Figure~\ref{fig:triangleHR_maggibands}.
In each panel the triangular region denotes the allowed HR values for positive count measurements. The top-right panel depicts the simulated HRs for a low M33 absorption (N$_{H}$ = 1$\times$10$^{20}$ cm$^{-2}$) from low temperatures ($\sim$~0.1~keV, smaller crosses) to high temperatures ($\sim$ 1~keV, larger crosses) at a range of abundance ratios. Points are color-coded based on the abundance ratio of O/Fe: blue crosses are those SNRs with high O/Fe ratios, indicating O enrichment, grey crosses are those SNRs for which the O/Fe ratio is near unity, and red crosses are SNRs for which the O/Fe ratio is low and thus indicative Fe enrichment. The unfilled circles are the 15 SNRs for which we were able to perform detailed spectral fits (Table~\ref{tab:fittab}), and are also color-coded based on their fitted abundance ratios. The bottom-left panel displays the simulated HRs for an intermediate M33 absorption value (N$_{H}$ = 1.2$\times$10$^{21}$ cm$^{-2}$), while the bottom-right has a high M33 absorption value (N$_{H}$ = 3.5$\times$10$^{21}$ cm$^{-2}$), both with the same spread of temperatures and abundance ratios as the top-right panel. The size of the points (both simulated and fitted) indicates temperature, with smaller crosses having lower temperatures. 

There is a clear trend with temperature in the simulated SNR HRs, wherein SNRs at a given O/Fe and N$_{H}$ move to the left and down (softer HRs) in Figure~\ref{fig:triangleHR_maggibands} as their temperatures go from high ($\sim$ 1~keV) to low ($\sim$~0.1~keV). This is denoted by the black arrow labeled ``kT" on each panel. The progression of the panels illustrates the changes to simulated HRs with changing absorption values, with increasing absorption moving SNRs at a given temperature and abundance ratio to generally harder HRs (up and to the right). At a given temperature and value of N$_{H}$ the O/Fe abundance ratio can move the HR diagonally downwards on the plot, as indicated by the black arrow labeled ``O/Fe". Some separation between abundance ratios is evident in the simulated sample, with SNRs at low O/Fe exhibiting larger HR1 values, possibly indicative of their stronger Fe L-shell lines and thus a Type Ia progenitor. 

The fitted sample (unfilled circles) is more consistent with the simulated sample at low N$_{H}$ (N$_{H}$ = 1$\times$10$^{20}$ cm$^{-2}$) for the majority of the fitted SNRs. The only SNR with low O/Fe in our fitted sample (red unfilled circle) is roughly consistent with the simulated sample assuming a lower temperature. There is one outlier in the fitted sample that does not clearly follow any of the simulated trends. This SNR has an intermediate abundance ratio value (grey unfilled circle) and is separated from the bulk of the population with the largest HR2 value. This is source XMM-132 (L10-081, LL14-119), and is classified as a CC SNR by LL14, and by \citet{Jennings2014} with a derived progenitor mass of $\sim$14M$_{\odot}$. 

We next look for correlations between SNR progenitor type and HR in our sample of 3$\sigma$ detections by cross-correlating our sources with those
from LL14 and \citet{Jennings2014} and assigning to each SNR, when available, a possible progenitor type based on their analyses of the surrounding stellar population. We illustrate this comparison between HRs and tentative progenitor type in the top-left panel of Figure~~\ref{fig:triangleHR_maggibands}, with HRs and their associated errors calculated from counts in the above bands using BEHR \citep{Park2006}. Each point is color-coded based on potential progenitor type: blue crosses represent potential CC classifications (nearby OB stars found in LL14), black squares are sources with CC classification from \citet{Jennings2014}, gray x's are sources for which there is no counterpart in LL14 or \citet{Jennings2014}, and filled red circles are sources with potential Type Ia progenitors (no nearby OB stars found in LL14). More transparent points have lower signal-to-noise ratios. The typical HR errors for sources with 1000 counts, 300 counts, and 100 counts are displayed for reference. We find no correlation between HR in these bands and potential SNR progenitor type. 

Ultimately we are far from the idealized case in the other panels of Figure~\ref{fig:triangleHR_maggibands}, and the range of temperatures, column densities, and abundance ratios probed, coupled with uncertainties on the HRs, do not allow for any kind of quantitative separation for SNR progenitor type based on HRs alone. In addition, it is perhaps not surprising to find no strong separation given that we expect only 17\% of the SNRs in M33 to be of Type Ia origin \citep{Mannucci2005}, which is about 18 total out of all 3$\sigma$ detections. 

It is also true that some Type Ia SNRs in the Large Magellenic Cloud (LMC) are Balmer-dominated, with little to no enhancement of [SII] emission \citep{Hughes1995}. As noted by \citet{Tuohy1982} this effect arises due to a fast shock propagating into a region of neutral hydrogen and giving rise to strong Balmer emission, while [SII] emission is suppressed in the high temperature region behind the shock due to low collisional rates. This implies that we may be missing the sample of young, ejecta dominated Type Ia SNRs in M33 by selecting SNRs mainly through their enhanced [SII]/H$\alpha$ ratios; however, we expect the population of young, Balmer-dominated Type Ia SNRs in particular to be quite small, as only 4 are reported in the LMC \citep{Tuohy1982,Hughes1995,Ghav2007,Maggi2016}. For older SNRs, the ejecta will be well-mixed with the surrounding circumstellar material, so evidence of the progenitor's ejecta signature would be diluted or erased. 

In addition to the bands defined by \citet{Maggi2014} we also test the correlation between potential SNR progenitor type (derived from LL14) and 
HRs based on counts in the $\textless$ 2~keV bands. This particular set of HRs was developed to take advantage of the 
soft-sensitivity of {\it XMM-Newton}. The ratios are calculated as follows: 

\begin{equation}
\begin{split}
HR1_{XMM} = (M_{XMM} - S_{XMM}) / (M_{XMM} + S_{XMM})\\
HR2_{XMM} = (H_{XMM} - M_{XMM})/ (H_{XMM} + M_{XMM})
\end{split}
\end{equation}

Here, the soft band is defined as 0.2-0.5~keV, the medium band is 0.5-1~keV, and the hard band is 1.0-2.0~keV. As outlined in W15 we used these HRs to isolate new SNR candidates 
based on the HR cuts described by \cite{Pietsch2004}, which are designed to take advantage of {\it XMM-Newton's} soft sensitivity. This method, combined with visual inspection
of the SNR candidates in [S II] and H$\alpha$, yielded the discovery of three new SNRs in M33 (first reported in W15). As before, we first compute the HRs in these bands based on SNR spectra simulated in {\tt XSPEC} with the model outlined above. The HRs computed from these simulated spectra are plotted with the same color-scheme as before in the top-right (N$_{H}$ = 1$\times$10$^{20}$ cm$^{-2}$), bottom-left (N$_{H}$ = 1.2$\times$10$^{21}$ cm$^{-2}$), and bottom-right (N$_{H}$ = 3.5$\times$10$^{21}$ cm$^{-2}$) panels of Figure~\ref{fig:boxHR_xmmbands}. The thick black arrow indicates the direction of increasing temperature (smaller to larger crosses). There is no clear distinction between abundance ratio values in these bands based on simulated spectra. The top-left panel of Figure~\ref{fig:boxHR_xmmbands} displays the HRs calculated from source counts in the above bands using BEHR \citep{Park2006}. Typical errors from BEHR for sources with 1000 counts, 300 counts, and 100 counts are displayed for reference. The points are again color-coded based on potential progenitor type from \citet{Jennings2014} or LL14. Similarly to the simulated data, there is no separation by progenitor type based on HRs in these bands. 

\subsection{X-ray Morphology: Power-Ratios}\label{sec:pr}

A basic question about an SNR is the nature of the supernova explosion. One way to tackle this question, as demonstrated by \citet{Lopez2009,Lopez2011}, is through the X-ray morphology of the SNR. Specifically, \citet{Lopez2009,Lopez2011} showed that the X-ray morphologies of young, ejecta-dominated SNRs are correlated with SN progenitor type as determined from other methods, like spectral fits. \citet{Lopez2009,Lopez2011} determined progenitor type (Type Ia versus CC) for a subsample of Milky Way and Magellenic Cloud SNRs through a multipole expansion of the X-ray surface brightness of each source. This method produces quantitative measurements of morphological asymmetry for SNRs, and is referred to as the ``power-ratio" method. \citet{Lopez2009, Lopez2011} find that for ejecta dominated SNRs, Type Ia SNe are ``statistically more spherical and mirror symmetric" than CC SNe, particularly in the 0.5-2.1~keV band.

Because this method has thus far only been applied to relatively nearby SNRs, we have performed a series of tests on a subset of the \citet{Lopez2011} data to determine the spatial resolution and number of counts necessary for determining SN progenitor type via the power-ratio method at distances greater than the Magellenic Clouds. To test the spatial resolution limits, we bin the data for a subsample of LMC SNRs sequentially until the resulting values change the quantitative morphologies. With each binning we recalculate the SNR 
centroid based on the new image. We find that binning the data by four, eight, and sixteen and recalculating the power-ratios preserves the separation between the two types. Our tests reveal that the decreased spatial resolution at the distance of M33 should 
not affect our ability to type SNRs based on morphology as long as the SNRs possess enough counts. However, at high enough binning, it becomes apparent that there are too few pixels to extract robust morphological information. The results become significantly unreliable when all the counts are contained in less than roughly 100 pixels, depending on number of counts. Thus the maximum distance at which we can apply this technique depends on the size of the SNR as well as the distance. We find a limiting distance for this method of $\sim$ 1200~kpc for the 0.5" {\it Chandra} pixel size and the largest SNR radius in \citet{Lopez2011} (r $\sim$ 30~pc). Adopting more conservative radii for young, ejecta-dominated SNRs of 20~pc and 10~pc yields limiting distances of $\sim$ 830~kpc, and $\sim$ 410~kpc, respectively. 

To test the count threshold, we take a random sampling of between 1 and 10\% of 
the original counts from a sample of LMC SNRs keeping the
images at full resolution and recalculating the power-ratios. We find that the method produces reliable results down to 2$\times$10$^{3}$ counts for Type Ia and down to 3$\times$10$^{2}$-4$\times$10$^{2}$ counts for CC SNe. Below these count thresholds the method begins to produce unphysical results, i.e. power ratios with errors that include negative values. CC SNe are more robust to this effect because they initially have higher power ratios for both the octopole and quadrupole moments; SNRs that have lower power-ratios with larger error bars require more counts to get a robust typing. Therefore, we find that at least  2$\times$10$^{3}$ counts are needed to robustly separate CC SNR progenitors from Type 
Ia SNR progenitors. Taken collectively, our tests demonstrate that SNRs with a radius of 20~pc and $\geq$ 2$\times$ 10$^{3}$ counts can have their progenitors typed via the power-ratio method out to the distance of M33. While there are a handful of SNRs in M33 with the requisite number of counts, none of these are large enough (r $\textgreater$ 20~pc) to utilize the power-ratio method for robust typing. We have verified this by testing the method on a few of the largest and brightest SNRs from the ChASeM33 survey and finding unphysical answers. In order to determine the quantitative morphologies of a large enough sample ($\sim$ 50) of SNRs in M33, our analysis suggests we would need an X-ray telescope with $\sim$ 0.03" resolution (17$\times$ the resolving power of {\it Chandra}), and with a 0.4 m$^{2}$ collecting area (10$\times$ the collecting area of {\it Chandra}).

\begin{figure*}
 \centering
    \begin{subfigure}
        \centering
        \includegraphics[trim=0 20 0 70,clip,scale=0.38]{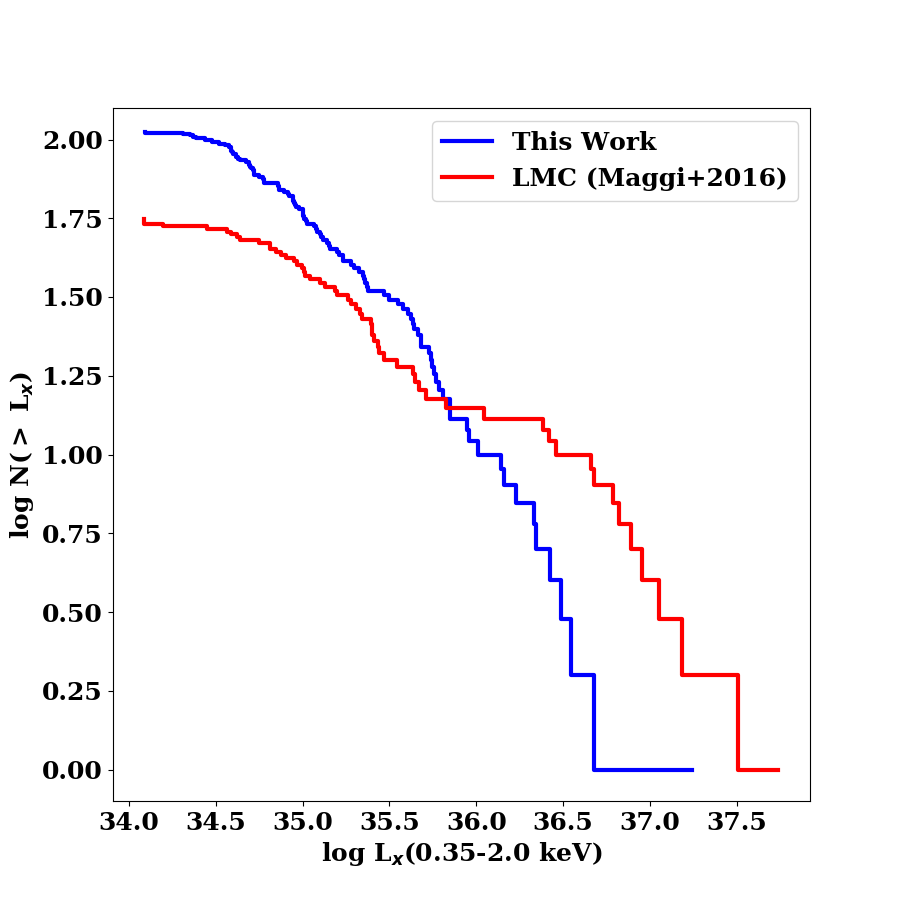}
    \end{subfigure}
    \begin{subfigure}
        \centering
        \includegraphics[trim=0 20 0 70,clip,scale=0.38]{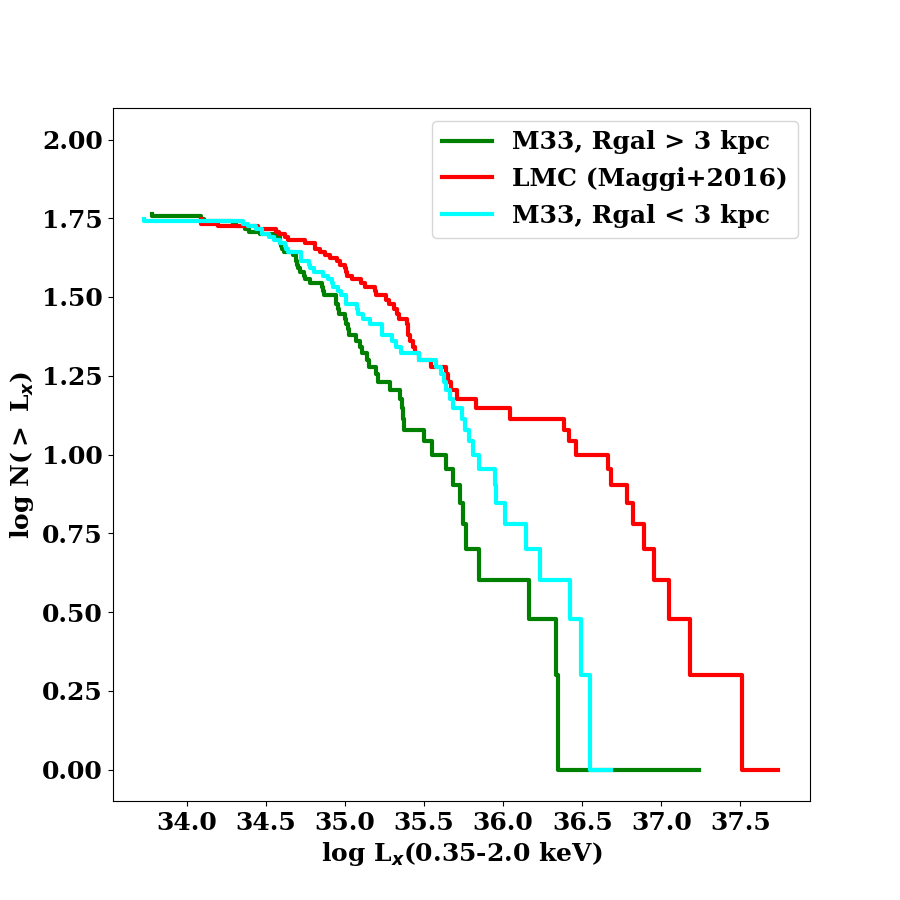}
    \end{subfigure}
\caption{{\it Left}: Cumulative X-ray luminosity function (XLF) for all 3$\sigma$ SNR detections from this work in blue. Cumulative XLF for the LMC from \citet{Maggi2016} in red. {\it Right}: Cumulative XLF for the inner 3 kpc of M33 (cyan, solar metallicity) and the outer 3 kpc (green, LMC-like metallicity) with the LMC XLF (red) for reference.}
\label{fig:snrxlf}
\end{figure*}

\section{Discussion}\label{sec:discuss}

Our deep XMM-Newton survey of M33 complements the high spatial resolution of the SNR candidates measured by the ChASeM33 survey with increased counts for spectral fitting, expanded survey area, and increased soft sensitivity for SNR detection. In this section we explore the implications of our results for the SNR X-ray luminosity function, and the X-ray detectability of SNRs.

\subsection{Supernova Remnant X-ray Luminosity Function} \label{sec:xlfsect}

We first construct the X-ray luminosity function (XLF) in the 0.35-2.0~keV band for all detections in this work (3$\sigma$ measurements) as shown in Figure~\ref{fig:snrxlf}. We overplot the XLF from \citet{Maggi2016} in red for comparison, transforming their 0.3-8.0 keV luminosities into the 0.35-2.0 keV band using WebPIMMS, and assuming an {\tt apec} spectrum with kT=0.6 keV, M33 hydrogen column density of of 1$\times$10$^{21}$ cm$^{-2}$, and a galactic absorption component of  6$\times$10$^{20}$ cm$^{-2}$. This SNR catalog has 11 detections with L$_{x}$ $\textgreater$ 10$^{36}$~erg~s$^{-1}$, as compared to 13 in the LMC, but fewer sources (three) at luminosities greater than $\textgreater$ 10$^{36.5}$~erg~s$^{-1}$ than are found in the LMC (eight). The limiting luminosities for this survey and that of \citet{Maggi2016} are of the same magnitude: L$_{x}$(0.2-2.0 keV) = 7.2$\times$10$^{33}$ ergs s$^{-1}$, and  L$_{x}$(0.3-8.0 keV) = 7.0$\times$10$^{33}$ ergs s$^{-1}$, respectively.

The shape of the LMC XLF, as discussed by \citet{Maggi2016} is clearly complex, and differs from the simple power-law distribution that can
be used to describe the M33 XLF. At the faintest end, there are likely incompleteness effects for both catalogs, but such claims cannot be made at the bright end, thus necessitating an explanation of the discrepancies for the population of bright SNRs.  We discuss several possible explanation below.

The explanation is unlikely to be the current star formation rate (SFR).  The SFRs of the two galaxies are similar: between 0.2--0.4 M$_{\odot}$~yr$^{-1}$ in the LMC over the last 100 Myr, with an increase to a rate of  0.4 M$_{\odot}$~yr$^{-1}$ occurring in the last 12 Myr  \citep{Harris2009}, and an average rate of 0.3 M$_{\odot}$~yr$^{-1}$ in M33 over the last 100 Myr \citep{Williams2013}.  Given these SFRs, both galaxies would be assumed to have close to the same rate of CC SNe production. 

Another possible explanation, as discussed by \citet{Maggi2016} is metallicity effects. In
particular, a lower metallicity environment will host stars with weaker line-driven
stellar winds. The consequence is smaller wind-blown cavities for massive stars such
that the SN explosion is running into a dense shell of material earlier in its evolution,
leading to brighter SNRs at earlier times \citep{Dwarkadas2005}. It is clear that the LMC
has more SNRs at the bright end than M33, but to test whether this is solely a
metallicity effect with respect to M33 one needs to take into account the metallicity
gradient in M33. To do so we construct the XLF for SNRs in M33 that are within
3 kpc of the galactic center, and the XLF for SNRs that are at galactocentric radii
larger than 3 kpc, as the metallicity in M33 goes from near-solar values within 3 kpc
to LMC-like metallicity outside 3 kpc \citep{Magrini2007}. The two M33 XLFs are depicted in the right-hand panel of Figure~\ref{fig:snrxlf}. In M33 the SNRs at higher metallicity ($\textless$ 3 kpc, cyan curve) have higher luminosities than those at lower metallicity ($\textgreater$ 3 kpc, green curve), which is exactly the opposite of the expected behavior if the luminosity differences are due to differences in progenitor wind mass-loss rates alone. If metallicity was the primary driver of differences in the SNR XLF one might expect the SMC, as the lowest metallicity galaxy, to host even more SNRs at the bright end than the LMC. As noted in \citet{Maggi2016} this is not the case. Furthermore, at later times the effects of SNe exploding into environments of differing densities would be largely erased, so metallicity effects on the surrounding medium would only be distinguishable for a younger population of SNRs.

\citet{Maggi2016} also found that SN type and ISM structure did not seem to play a strong role in the SNR XLF shape in the LMC.  The ratio of Type Ia versus CC SNRs in the XLF is difficult to compare across galaxies, as we do not have a definitive set of SN types for a large sample of M33 SNRs; however, both galaxies are likely dominated by core-collapse events. \citet{Maggi2016} note that the ratio of CC to Type Ia SNRs is slightly higher at the bright end of the LMC XLF, but not strongly so.  Furthermore, while differences in the SNR brightness distribution could also arise from SNe that are exploding into non-uniform interstellar medium, \citet{Maggi2016} found no significant spatial correlation between SNRs in different luminosity bins and HI maps of the LMC. However, HI maps may be an imperfect indicator of ``local" density around an SNR, so we cannot fully rule out that local density variations in the vicinity of SNRs contribute to different SNR luminosity distributions. In fact, one may even expect a more uniform ISM in a large, spiral galaxy like M33, as compared to LMC, which would result in SNRs with lower X-ray luminosities in an M33-like galaxy. 

Finally, it may be that the most plausible explanation for the differences in SNR XLF shapes is that the star formation histories (SFHs) are different on a 50 Myr timescale, which would be relevant for SNe production.  The total LMC SFR
has increased by a factor of 2 over the past 50 Myr \citep{Harris2009} which could result in a top-heavy progenitor mass distribution compared with a constant SFR.  For example, SN1987A had a relatively massive progenitor (20 M${\odot}$; \cite{woosley1988}).  We do not possess a global and resolved star formation history for M33, but based on the SNR progenitor mass distribution in M33 with peak mass at around 8 M$_{\odot}$ from \citet{Jennings2014} it is likely that M33 has a near constant SFR on this timescale.   Furthermore, M33 is relatively high-mass and isolated compared to the LMC, making it less likely to change its global SFR significantly on timescales as short as 50 Myr.  Thus, it is possible that differences in the XLF distributions at the bright end between the LMC and M33 are due the progenitor mass distributions leading to more bright, young SNRs in the LMC than in M33.

\begin{figure}
\centering
\includegraphics[trim=0 0 0 50,clip,scale=0.4]{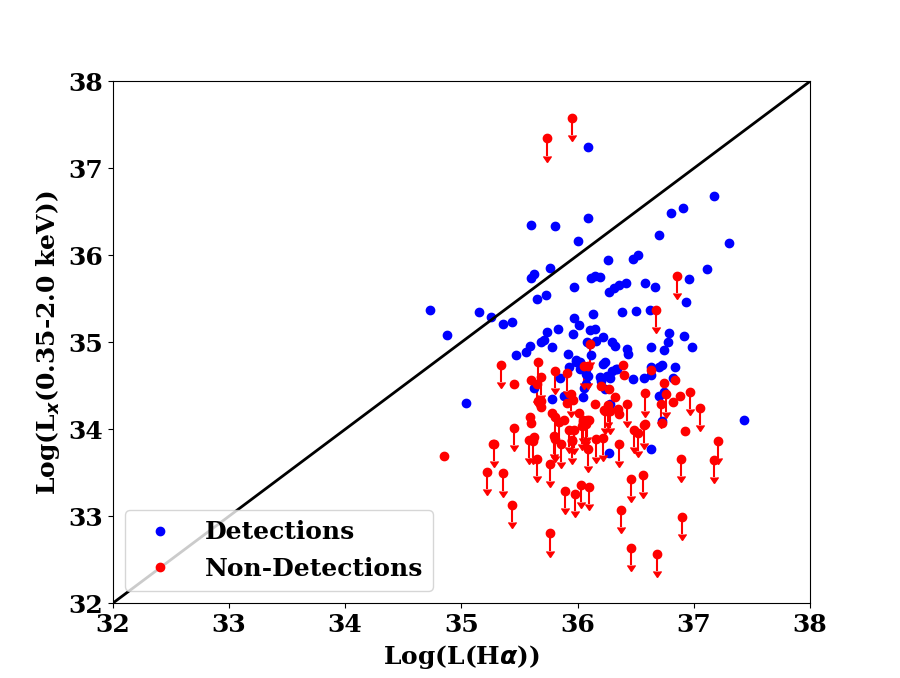}
\caption{Comparison between the luminosities in 0.35-2.0~keV band from this work and the H$\alpha$ luminosities from LL14. Red circles are SNR candidates non-detected in X-rays, and blue circles are SNRs detected at the 3$\sigma$ level in X-rays. Sources that lie to the left and above the black diagonal line have X-ray luminosities greater than their H$\alpha$ luminosities.}
\label{fig:lumcompare}
\end{figure}

\begin{figure*}
\centering
\includegraphics[trim=60 0 50 0,clip,scale=0.47]{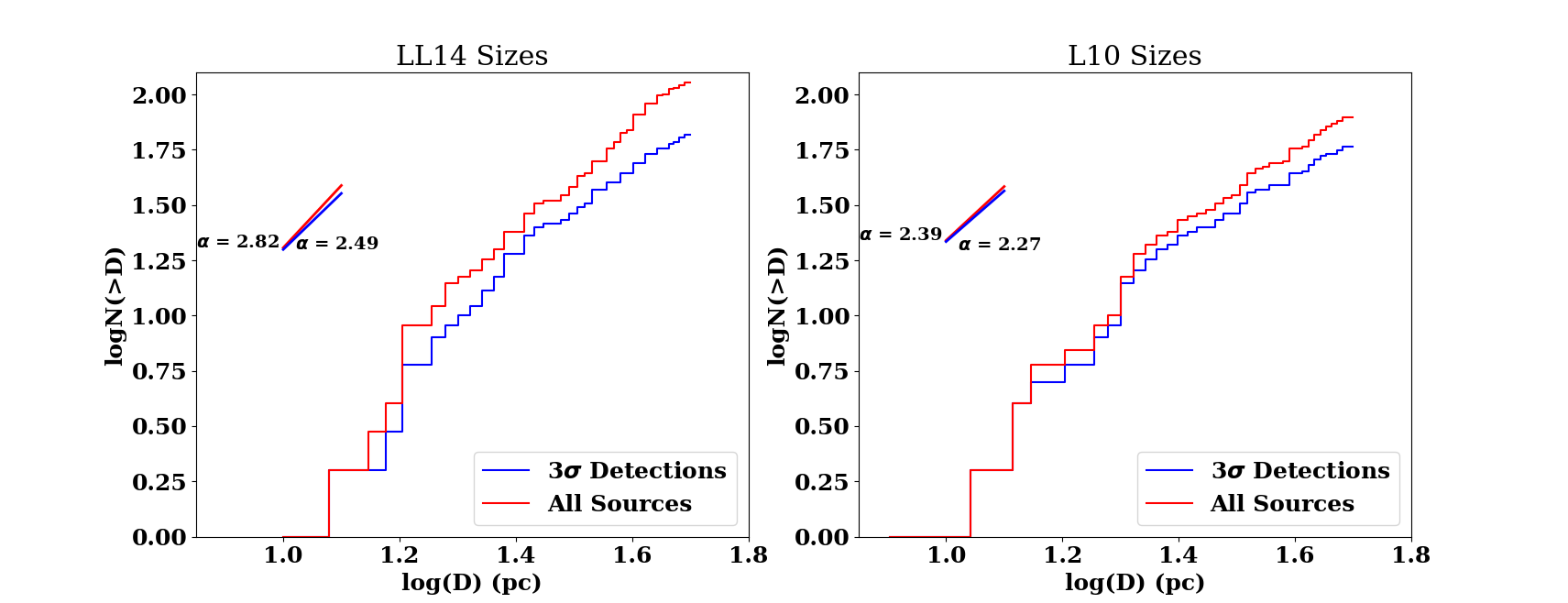}
\caption{Cumulative size distribution for all 3$\sigma$ detections (blue) versus all candidates (red) with D $\textless$ 50~pc in the sample for sizes measured by LL14 (left panel) and sizes measured by L10 (right panel). We measure slopes of $\alpha$ $\sim$ 2.8 and $\alpha$ $\sim$ 2.5 for all sources and all detections, respectively, using LL14 sizes. We find slopes of $\alpha$ $\sim$ 2.4 and $\alpha$ $\sim$ 2.3 for all sources and all detections, respectively, using L10 sizes. The slopes for all detections are in good agreement with the slope of $\alpha$ = 2.5 expected of a population of SNRs in the Sedov phase.}
 \label{fig:nlessd}
\end{figure*}

\subsection{Detectability}\label{sec:detect}

The M33 XLF appears to flatten around 3$\times$10$^{34}$~erg~s$^{-1}$, implying that our sample may still be incomplete at the faintest luminosities, and that with increased sensitivity the entire SNR population of M33 could be detected. To explore X-ray detectability we first compare the luminosities in the 0.35-2.0~keV band for all sources to the H$\alpha$ luminosities from LL14. In Figure~\ref{fig:lumcompare} we plot the X-ray luminosity in this band versus the H$\alpha$ luminosity and find no significant correlation
between the luminosities for either X-ray detections (blue points) or X-ray non-detections (red points, sources from the optical catalogs that were measured at the 2$\sigma$ or upper-limit level in X-rays). Similarly to L10, we find that only one SNR has an X-ray luminosity significantly higher than its H$\alpha$ luminosity (i.e. falls above the black line), and this is the brightest SNR in the sample, XMM-041 (L10-025). While it does appear that SNRs that are brighter at X-ray wavelengths also have generally higher H$\alpha$ luminosities, there is a large spread in the X-ray to H$\alpha$ luminosity comparison. The lack of strong correlation between luminosities can be explained by differences in the regions of the SNR being probed by each diagnostic. Namely, the X-ray luminosity is probing the region of the reverse shock, which is generally a region of higher temperature that cools more slowly. The H$\alpha$ luminosity, by contrast, originates from recombination in the cooler, more dense shell region, which tends to cool on shorter timescales \citep{LongSNR, Leonidaki2013}. The lack of correlation may also be due to the presence of non-uniform ISM, rather than regions of differing temperatures \citep{Pannuti2007}.

\begin{figure}
\centering
\includegraphics[trim=30 30 0 70,clip,scale=0.4]{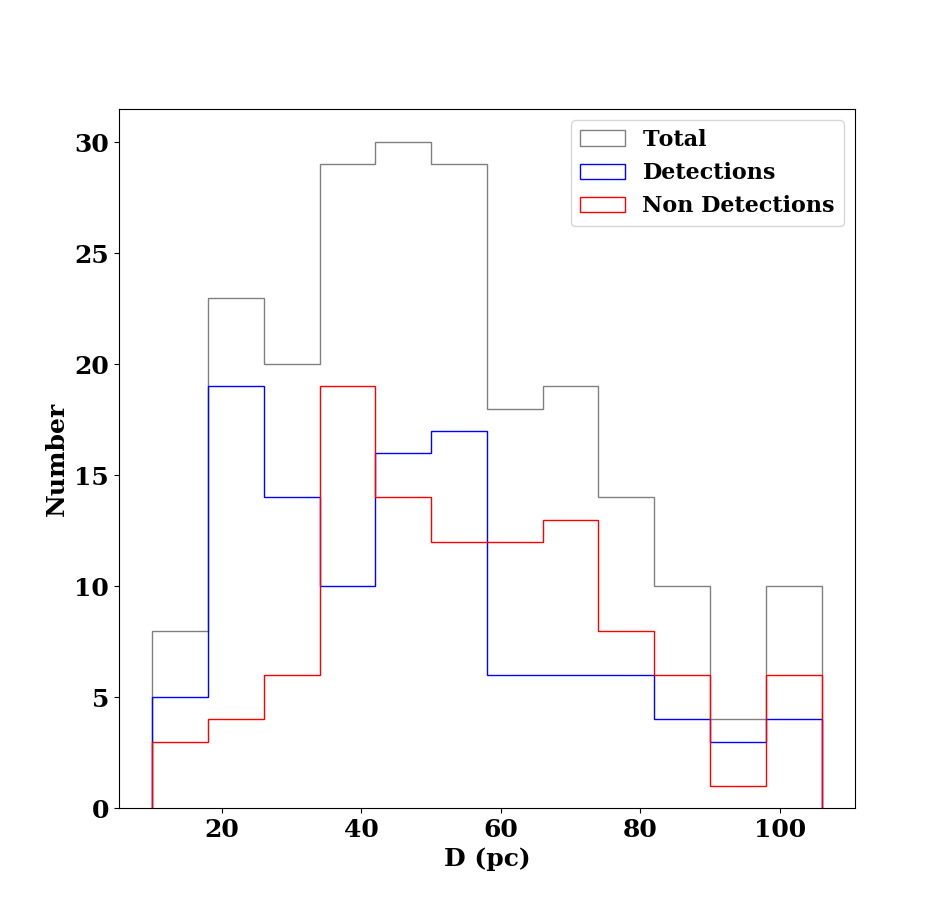}
\caption{Histogram of sizes from L10 and LL14 for all non-detections (red) and detections (blue) in this sample. All sources with D$\textgreater$~100~pc are put into the rightmost bin.}
 \label{fig:sizehist}
\end{figure}

\begin{figure*}
    \centering
    \begin{subfigure}
        \centering
        \includegraphics[trim=20 20 20 0,clip,scale=0.35]{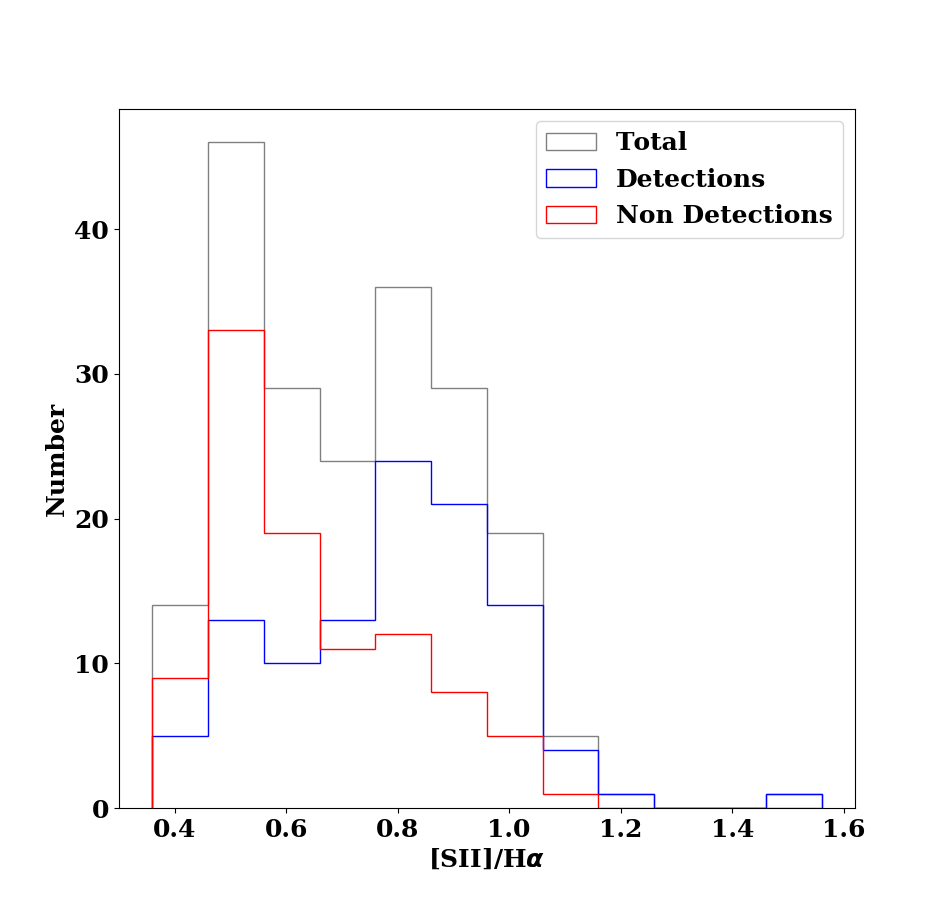}
    \end{subfigure}
    \begin{subfigure}
        \centering
        \includegraphics[trim=20 20 20 0,clip,scale=0.35]{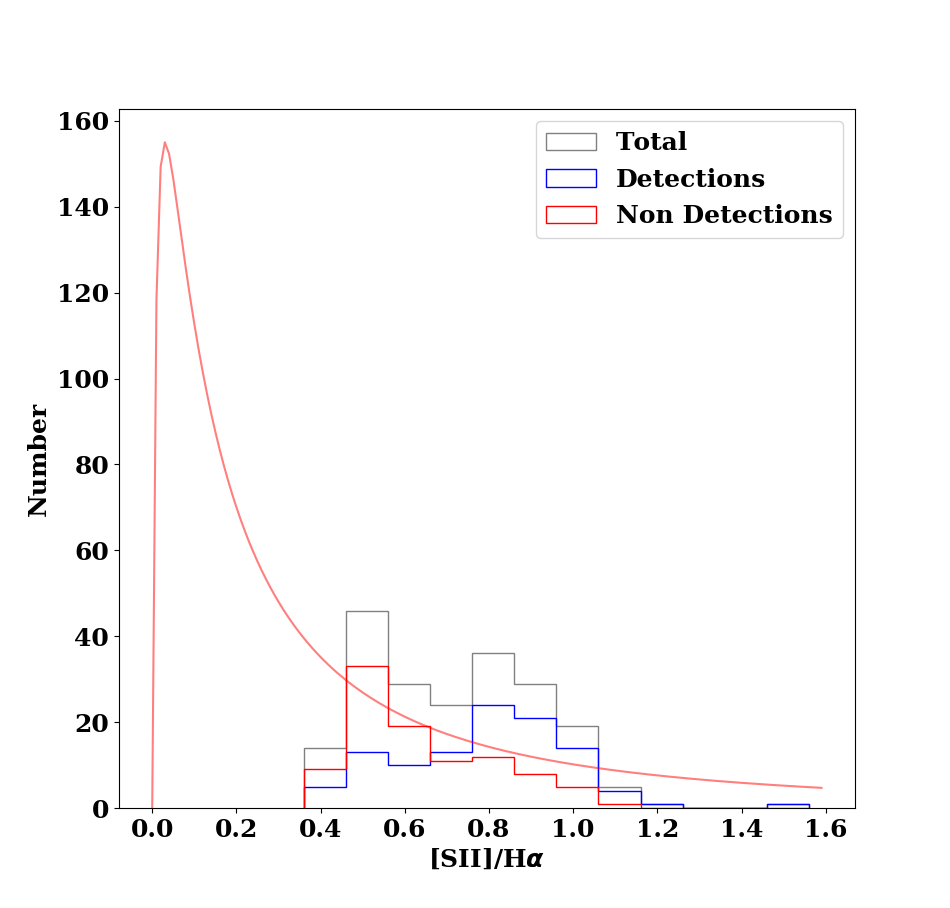}
    \end{subfigure}
     \begin{subfigure}
        \centering
        \includegraphics[trim=0 20 20 0,clip,scale=0.35]{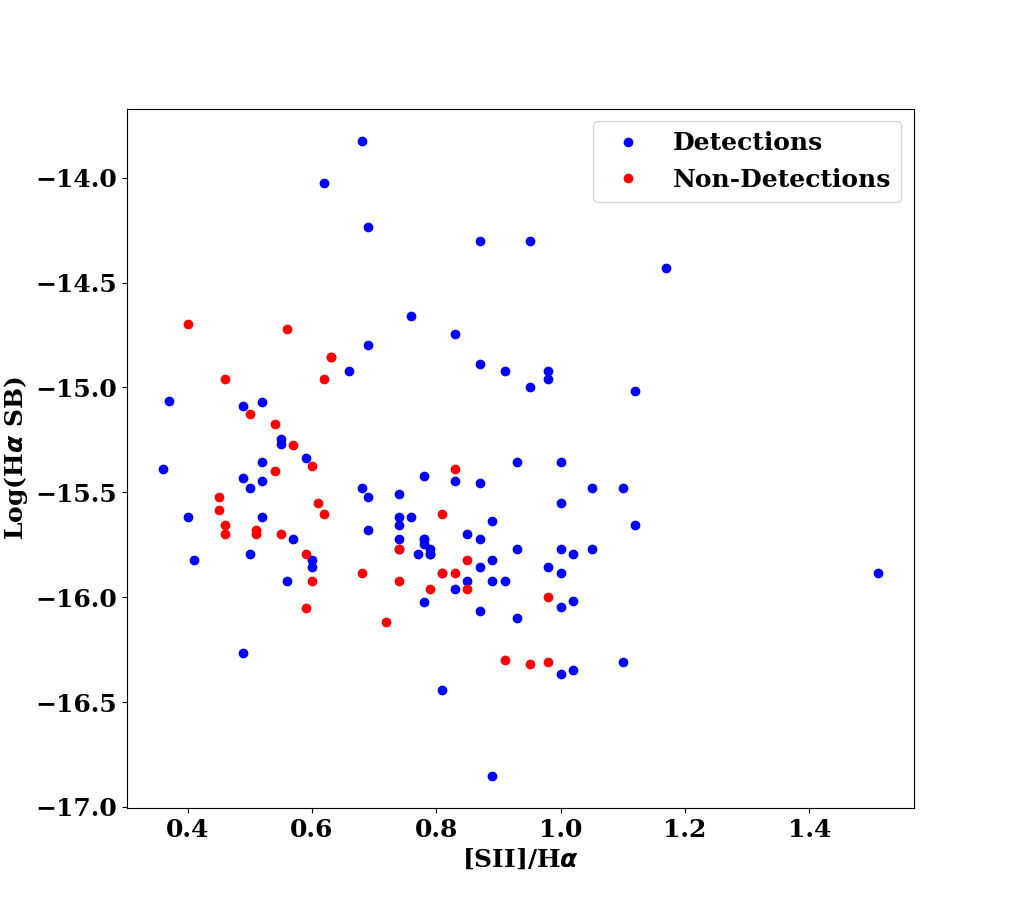}
    \end{subfigure}
    \begin{subfigure}
        \centering
        \includegraphics[trim=20 20 20 0,clip,scale=0.36]{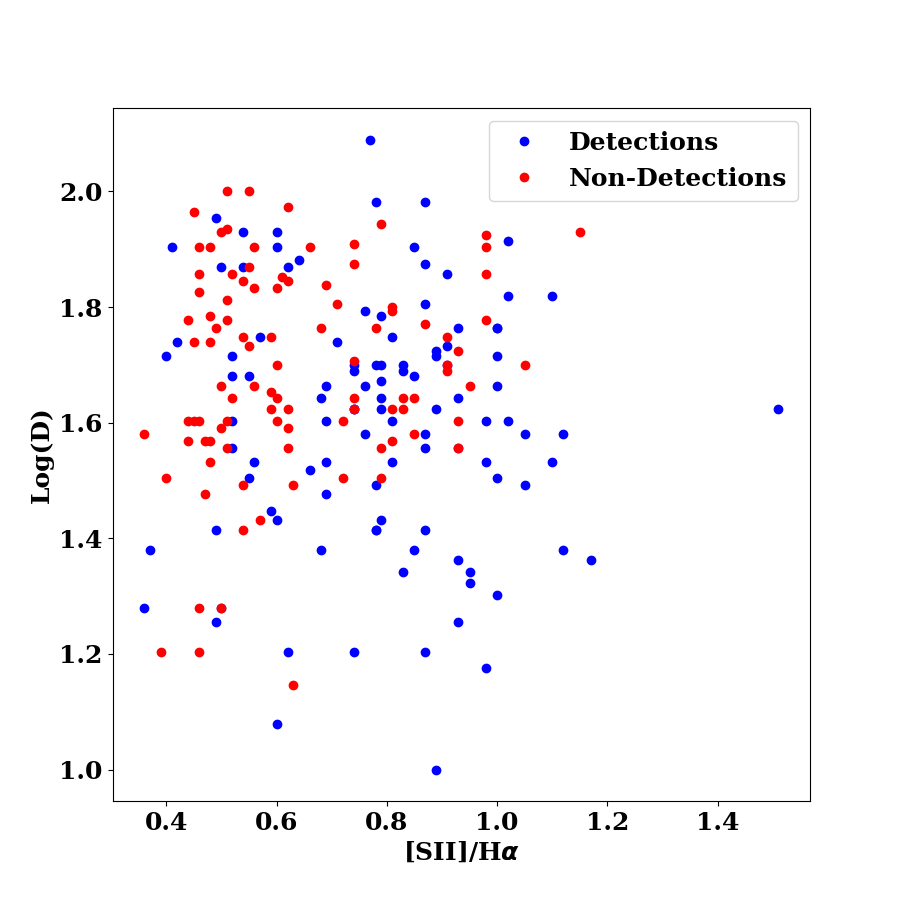}
    \end{subfigure}
 \caption{{\it Top left}: Histogram of [SII]/H$\alpha$ ratios from L10 and LL14 for all SNR candidates non-detected in X-rays (red) versus X-ray detected SNRs (blue) in the sample.  {\it Top right}: A log-normal error distribution of line ratios with a mean of 0.1 and $\sigma$ value of 0.1 overplotted with respect to the population of X-ray non-detections. {\it Bottom left}: [SII]/H$\alpha$ ratios from L10 and LL14 for all SNR candidates non-detected in X-rays (red) versus X-ray detected SNRs (blue) versus the measured H$\alpha$ surface brightness values from L10. {\it Bottom right}: [SII]/H$\alpha$ ratios from L10 and LL14 for all SNR candidates non-detected in X-rays (red) versus X-ray detected SNRs (blue) versus SNR diameters in pc.}
\label{fig:siihahist}
\end{figure*}

\begin{figure}
\centering
\includegraphics[trim=0 20 0 0,clip,scale=0.4]{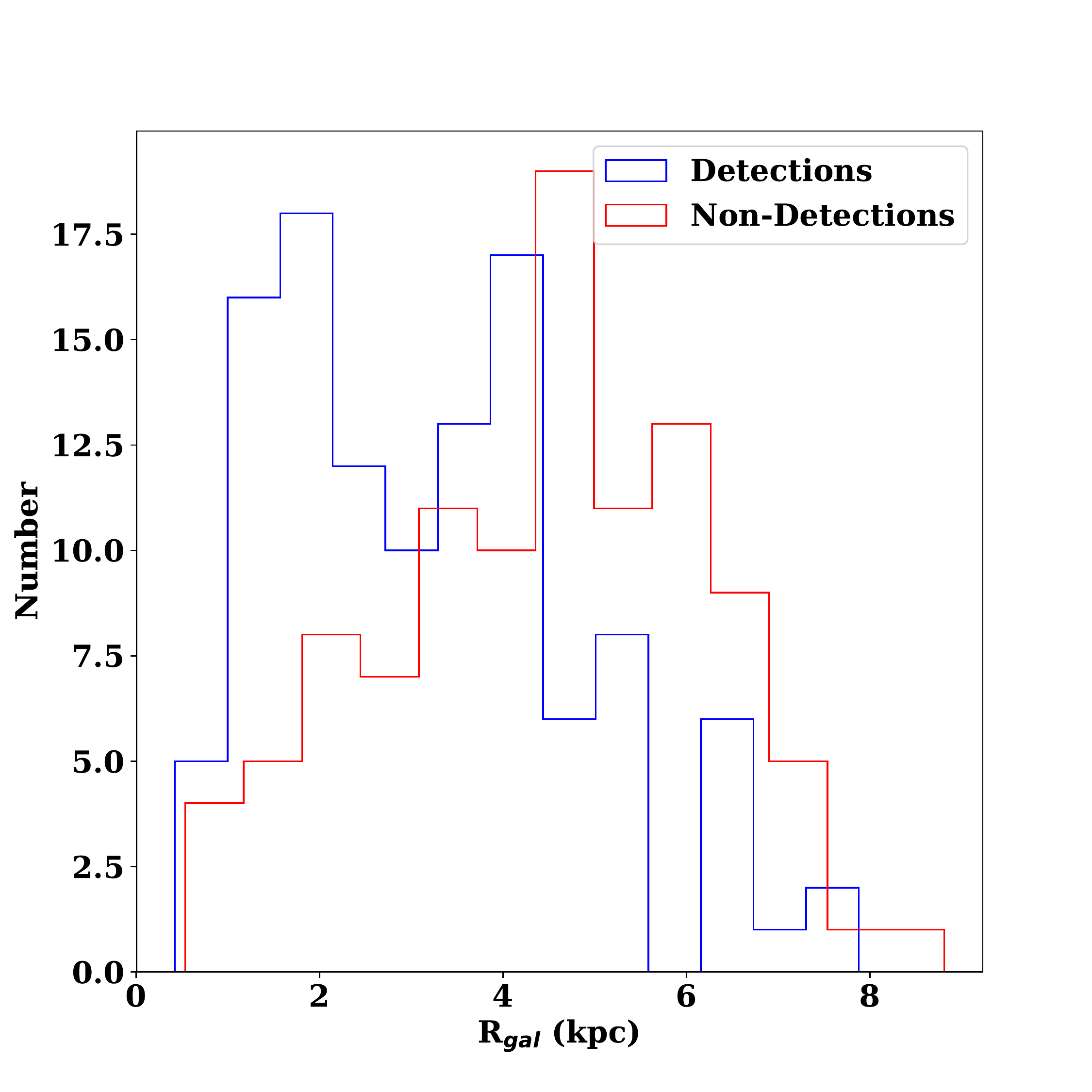}
\caption{Histogram of galactocentric radii for all X-ray non-detections (red) and X-ray detections (blue) in this sample.}
 \label{fig:rgalhist}
\end{figure}

We also compare the cumulative size distribution for all X-ray detected SNRs versus all sources (detections and candidates) in our catalog with D $\textless$ 50~pc to look for differences between the slopes of the distributions of each population. We choose this size cutoff, because the population of SNRs and SNR candidates is not complete above D $\sim$~50~pc. Our results are plotted in Figure~\ref{fig:nlessd} both for sizes from LL14 (left panel) and sizes from L10 (right panel). We find that the addition of 2$\sigma$ and upper-limit measurements to the cumulative distribution tends to steepen the slope. This is likely because the population of SNR candidates non-detected in X-rays are biased towards larger diameter sources, as can also be seen in Figure~\ref{fig:sizehist}. Likewise the slope of the cumulative distribution when using the LL14 sizes is steeper than the slope when using only L10 sizes, owing to the fact that the LL14 radii are systematically larger than those measured by L10. The slopes of the cumulative distributions for the 3$\sigma$ SNR detections only are $\alpha$ $\sim$ 2.5 and $\alpha$ $\sim$ 2.3 when using LL14 and L10 size measurements, respectively. Both measured slopes are consistent with $\alpha$ = 2.5, which is the slope expected for a population of SNRs in the Sedov phase. However, there are also various selection effects that can lead to biases in the sample of SNRs in a cumulative size distribution. For example, ISM conditions can strongly affect both the size and luminosity evolution of an SNR, though without detailed constraints on local ISM conditions we are unable to quantify the impact of such an effect. Surveys such as this one that confirm SNR candidates primarily on the basis of both optical and thermal X-ray emission are also liable to miss some young, X-ray emitting SNRs, thus biasing an optically selected and X-ray confirmed sample towards larger diameter SNRs. 

In addition to the cumulative size distribution of the sample, we also look at the overall size distribution of all detections versus non-detections at all diameters. We plot this distribution in Figure~\ref{fig:sizehist}, with all sources at D $\textgreater$ 100~pc placed in the rightmost bin. The X-ray detected SNR sample extends to smaller diameters, implying that most small diameter SNRs are detected in X-rays. By contrast, the X-ray non-detections display a bias towards larger sizes, and have a sharper cutoff at smaller diameters than the population of sources detected in X-rays. The difference in diameters between the X-ray detected and X-ray non-detected sample may be attributed to age, or evolutionary effects, as young SNRs in the free expansion phase are likely to display X-ray emission, while older SNRs in the radiative phase show stronger optical emission \citep[e.g.][]{Leonidaki2010}. This difference in sizes between the two populations leads to a steepening of the slope of the cumulative distribution when SNR candidates that are non-detected in X-rays are included. 

Finally, we compare the distribution of [SII]/H$\alpha$ ratios for SNRs detected in X-rays versus those candidates undetected in X-rays in this catalog. The [SII]/H$\alpha$ ratio is typically used as a way to distinguish optical emission from shocked regions in SNRs from emission from HII regions, with a cutoff at $\textgreater$ 0.4 for classification as an SNR candidate. Higher [SII]/H$\alpha$ values are indicative of regions with radiative shocks where enough recombination has occurred to produce significant [SII] emission, as in SNRs. In Figure~\ref{fig:siihahist} we demonstrate that there are two distinct populations in the [SII]/H$\alpha$ distribution, with the SNR candidates non-detected in X-rays being drawn from a distribution with on average lower measured [SII]/H$\alpha$ than the population of X-ray detected SNRs. To determine if these populations are physically distinct, we look for correlations between [SII]/H$\alpha$ and object size and surface brightness as shown in the bottom two panels of Figure~\ref{fig:siihahist}. We find no strong correlation between [SII]/H$\alpha$ ratio and object size, and only a slight correlation between surface brightness and this ratio, with the lower surface brightness non-detections displaying on average higher [SII]/H$\alpha$ values (bottom-left panel of Figure~\ref{fig:siihahist}). It is possible that some of the X-ray non-detections at low [SII]/H$\alpha$ (but above the 0.4 SNR candidate threshold) could represent the tail-end of a log-normal error distribution of line ratios of ionized nebulae. An example of such a distribution containing $\sim$ 480 sources with a mean [SII]/H$\alpha$ value of 0.1 and $\sigma$ $\sim$ 0.1 is over plotted in red on the top-right panel of Figure~\ref{fig:siihahist}.

The application of the cutoff at [SII]/H$\alpha$ $\sim$ 0.4 for optically identifying SNR candidates should not be discounted based on a number of X-ray non-detected SNR candidates that also fall above this threshold, as this may be due to differences in varying shock conditions, or circumstellar environment on small scales. For example, SNRs that have not encountered enough dense material may not form radiative shocks, and would therefore not display high [SII]/H$\alpha$ values. Similarly, if the metallicity of the ISM is non-uniform one might expect different distributions of [SII]/H$\alpha$ depending on location in the galaxy. To test for environmental differences we construct a histogram of the galactocentric radii for all X-ray detected SNRs (blue) and X-ray non-detected SNR candidates (red) in Figure~\ref{fig:rgalhist}. The population of X-ray detections (sources with higher [SII]/H$\alpha$, as can be seen from the upper left plot of Figure~\ref{fig:siihahist}) are located at preferentially smaller galactocentric distances than the population non-detected in X-rays (sources with lower [SII]/H$\alpha$). We find no evidence that this separation is due to a gradient in exposure time or detector location in the observations. Instead, the separation may point to a metallicity effect, as there is a known chemical abundance gradient in M33, with the highest metallicities occurring at galaxy center and decreasing outwards \citep{Magrini2007, Neugent2014}. In particular, \citep{Magrini2007} measure this gradient as comprised of two slopes with the break occurring at R $\sim$ 3 kpc, similar to the radius at which we see the separation between the two histograms in Figure~\ref{fig:rgalhist}. Alternatively, this separation could be due to the effects of differing densities, with higher densities in the inner parts of the galaxy leading to stronger X-ray emission, and high [SII]/H$\alpha$ values.

\section{Conclusions}\label{sec:conclusions}

We have carried out a deep {\it XMM-Newton} Survey of M33 to
complement the one performed by {\it Chandra}.  With the power of both
datasets we have detected at 3$\sigma$ confidence $\approx$ 50\% of the SNR candidates in M33 from previous X-ray and optical surveys (e.g. L10, LL14). These 105 sources are all robust SNR detections verified by both optical and X-ray measurements. We performed detailed spectral fitting for 15 SNRs, twice the number possible than with the {\it Chandra} data
alone. We find evidence of elevated O/Fe values from X-ray spectral fits for one SNR (XMM-068), implying that this SNR exploded in a region generally enriched by CC ejecta. Based on the fitted spectral parameters we also determine that the majority of the brightest SNRs are old (t $\textgreater$ 1000 yrs), ISM-dominated SNRs. 

To complement the spectral fitting analysis we have also tested the ability to type SNRs based on HRs in custom energy bands and X-ray morphology. We conclude that HRs or colors alone are too coarse as methods for detailed typing due to uncertainties in HRs coupled with degeneracies between the lines contributing to specific energy bands, SNR temperatures, and absorption values. In addition, due to current limits on telescope collecting area and resolving power we are unable to distinguish the SN progenitor type for a large sample of SNRs in M33 using quantitative morphology. However, the combination of quantitative morphology with HRs for SNRs in the much nearer Magellenic Clouds yields promising results for typing extragalactic SNRs independent of detailed spectral analysis for all SNRs in a sample. 

We also use our large sample of SNRs to construct an XLF in both the inner ($\textless$ 3 kpc, solar-like metallicity), and outer ($\textgreater$ 3kpc, LMC-like metallicity) portions of M33 to test for metallicity effects on the luminosity distribution of the SNR population. In comparing XLFs in the inner and outer regions to one another, and also to the LMC SNR XLF, we find that while metallicity may play a role in SNR population characteristics, differing star formation histories on short timescales, and small-scale environmental effects appear to cause more significant differences between X-ray luminosity distributions.

Finally, we perform an analysis of the X-ray detectability of the M33 SNRs based on their physical properties. We compare this X-ray detected population of SNRs to the population of SNR candidates for which we have 2$\sigma$ or upper-limit measurements in X-rays. The latter population is larger in diameter, located at preferentially larger galactocentric radii, and has lower measured [SII]/H$\alpha$ values than the former. These differences suggest that the X-ray non-detected SNRs are likely comprised of a mixture of larger and/or fainter SNRs that potentially exploded into less dense, lower metallicity mediums that fall below our detection threshold, and some photoionized regions (HII regions or regions of diffuse ionized gas) whose measurement errors in the optical place them above the [SII]/H$\alpha$ ratio cut used by most surveys. If we include only the X-ray detected SNRs in the cumulative size distribution, the distribution has a slope of 2.5, in accordance with a population of SNRs in the Sedov phase of evolution.  

Future work will expand upon this large sample of well-characterized SNRs by exploring in more detail the interplay between host galaxy environmental factors and the resulting SNR properties. In particular, a more systematic study of surrounding ISM properties, coupled with resolved star formation histories in the vicinities of M33 SNRs will further quantify the dominant drivers behind SNR detectability, and add to the sample of SNRs with determined progenitor types.

Support for this work was provided by NASA grants
NNX12AD42G and NNX12AI52G. T.J.G. and P.P.P.
acknowledge support under NASA contract NAS8-03060 with
the Chandra X-ray Center. We thank the referee for their very helpful comments in improving the manuscript.

{\large \bibliography{references} }
\bibliographystyle{mnras}

\appendix

\section{Spectral Extraction Techniques}\label{extracttech}

The primary tool for spectral extraction is the {\tt SAS} task \textit{evselect}. One can use the selection expression of this task to define the location, in image coordinates, where the extraction
should take place. The nature of this survey allows for a particular source to be found in different
combinations of instruments and fields of data. FITS images were created with World Coordinate
Systems for each of the 30 brightest sources. Background regions were selected manually using
criteria recommended by the {\it XMM-Newton} Calibration Technical Notes. For the
EPIC-MOS instruments, either an annulus around the source or a separate location that has an
equivalent off-axis angle, related to the vignetting, and on the same CCD should be used. For
EPIC-pn, the background extraction region should not be an annulus due to the possibility of
out-of-time (OOT) events interfering. Backgrounds should instead be taken on the same CCD if
possible and at an equivalent readout distance on the CCD (the same RAWY value). Selecting
the background manually also allowed for the best possible location to be chosen, maximizing the
value of the spectra. Through examination of spectra using three differently sized background
regions--same extraction area as source, double the area, and ten times the area--we determined
the optimal background size to use was double the area of the source region. Any larger and
surrounding sources would make finding a source-free background difficult while trying to adhere
to the suggested parameters. Using a simple script, these paired region locations were saved to text
files in their observations specific image coordinates.

The \textit{evselect} selection expressions, along with some parameter values like bin size and maximum
channel, vary between the MOS and pn instruments. Depending on the instrument, the proper
image coordinates for the source were fed to \textit{evselect} and then repeated with the background
region detector coordinates. The source and background extraction regions of the spectra were computed, followed by the
generation of the redistribution matrix file (RMF) and ancillary response file (ARF). The RMF
and ARF file names were written to the RESPFILE and ANCRFILE header keywords of both the
source and background spectra using the {\tt HEASARC FTOOLS} software task \textit{grphha}.

If a source lay within the field of view of the PMH 47 observations (see W15 Section 2), we elected to combine those
spectra using the {\tt FTOOLS} software. Due to the varying roll angle of the PMH 47 observations,
it was possible for a source to be out of the field of view or on a chip gap for one or more of the
observations. This necessitated taking care to properly merge the header keywords, of which two
were critical. The EXPOSURE keyword is simply summed, but the BACKSCAL keyword, which
provides the number of sky pixels in the extraction area of the source was dealt with more carefully
when the effective area (ARF) and response matrix (RMF) files from the different observations
were combined. In particular, the BACKSCAL values(B$_{i}$) were weighted by the exposure time
(E$_{i}$) in the manner of Huenemoerder, Davis, Houck and Nowak (2011). 

\begin{gather*}
B_{final} = \frac{1}{E_{total}}\sum_{i} (B_{i}E_{i})
\end{gather*}

The source and background spectra were merged two at a time using {\tt mathpha} and without
error propagation as we decided to perform the error propagation based on counts alone. The final merged product had the BACKSCAL and summed EXPOSURE
keywords written to the header. The ARF and RMF files of each observation were first individually
combined to create a response file using {\tt mkarfrmf}, and were then weighted by exposure and merged
together using {\tt addrmf}. In order for the combined files to work correctly in {\tt XSPEC}, two additional
keywords, POISSERR and STAT ERR needed to be changed in the source and background spectral
files. When combining the spectra, the {\tt mathpha} task also created a STAT ERR column which was
found to be too conservative. This column was deleted and the updated keywords allowed error
propagation to be determined based on the counts alone.

\bsp	
\label{lastpage}
\end{document}